\documentclass[preprint,12pt]{elsarticle}

\usepackage{amsmath,amssymb,amsfonts}
\usepackage{graphicx}
\usepackage{cleveref} 
\usepackage[dvipsnames]{xcolor}
\usepackage{textcomp,nicefrac}
\usepackage{notoccite}
\usepackage{booktabs}
\usepackage{url}
\usepackage{etoolbox}

\journal{Instruments and Methods in Physics Research Section A: Accelerators, Spectrometers, Detectors and Associated Equipment}

\begin{document}

\begin{frontmatter}



\title{Pixelated Plastic Scintillator Array Manufacturing using Fast-, Photo-Curable Resin}


\cortext[*]{Corresponding authors.}   

\author[a]{C.J. Moore, \corref{*}}
\ead{chandler.moore.3@us.af.mil}
\author[a]{J.J. Manfredi, \corref{*}}
\ead{juan.manfredi@us.af.mil}
\author[a,b]{M. Febbraro,}
\author[a]{D. Rutstrom,}
\author[c]{A. Decker,}
\author[a]{R. Kemnitz,}
\author[b]{T. Ruland,}
\author[b]{P.A. Hausladen}

\affiliation[a]{organization={Air Force Institute of Technology},
            city={Wright-Patterson AFB},
            postcode={45433}, 
            state={OH},
            country={U.S.A.}}
\affiliation[b]{organization={Oak Ridge National Laboratory},
            city={Oak Ridge},
            postcode={37830}, 
            state={TN},
            country={U.S.A.}}
\affiliation[c]{organization={Defense Threat Reduction Agency Nuclear Science and Engineering Research Center},
            city={West Point},
            postcode={10996}, 
            state={NY},
            country={U.S.A.}}

\begin{abstract}
Pixelated plastic scintillator arrays can serve as high efficiency and high resolution neutron imaging detectors. Manufacturing these arrays is intensive in both time and labor. This work presents a fabrication method based on additive manufacturing for two-dimensional plastic organic scintillator arrays using a custom-built automated assembly machine and a custom photocurable resin that has significant non-aromatic acrylate oligomer content. The process involves two main stages: fully autonomous production of one-dimensional layered arrays, followed by semi-autonomous cutting and stacking to form two-dimensional pixel arrays. One-dimensional arrays were manufactured at a rate of around 4 layers per hour with minimal defects and tight dimensional tolerances, while two-dimensional arrays up to 7 $\times$ 7 pixels and 70 mm in length were completed in approximately 3.5 hours. Final arrays exhibited dimensional deviations of less than 0.5 mm. Two-dimensional arrays read out by a multi-anode photomultiplier tube demonstrated per-pixel position resolution and pulse-shape discrimination, enabling gamma–neutron interaction separation in mixed radiation environments.
\end{abstract}




\end{frontmatter}


\section{Introduction}
\label{sec:introduction}
Radiation imaging is useful for detection of diverted or illicit material, such as at border crossings \cite{Sowerby2009, Park2022, Radle2010}. The most commonly used imaging modality is X-ray imaging. However, X-rays suffer from limited penetration of lead and other heavy metals. Fast neutron imaging provides a potential alternative. Fast neutrons from compact monoenergetic sources offer better penetrability for high-Z materials and induce fast neutron interactions that can be used to identify both fissionable and hydrogenous substances \cite{Radle2010, Mihalczo2012, Wright2018, HEATH2024}. However, current systems that perform fast neutron imaging and material identification \cite{Cutmore2010, EBERHARDT2005} are limited by their present spatial resolution, efficiency, and the practical ability to manufacture detector geometries that are both efficient and substantially higher resolution.


Fast neutron detection in an imaging context is often accomplished via recoil protons in a plastic scintillator material. Plastic scintillators are common neutron detection materials because of their ruggedness, low cost, and ability to discriminate between neutrons and gamma rays \cite{Zaitseva2012}. The stopping distance of the recoil protons limits the detected interaction resolution and therefore the resolution of the neutron image \cite{bilheux}. Small features can be resolved at a stopping distance of 1-2 mm using 14.1 MeV neutrons, but this requires integrating detectors and prolonged exposure times \cite{bilheux, OKSUZ2022}. For monolithic integrating scintillators, increasing detector material thickness improves detection efficiency, but at the cost of spatial resolution. One way to recover fine spatial resolution while increasing thickness (and therefore efficiency) is to segment the scintillator into pixels using reflective material. This optical segmentation helps confine scintillation light within a given pixel, allowing accurate localization of interactions even in a thick detector. However, manufacturing pixelated arrays is more complex and time-consuming than producing a single scintillator block, with costs rising as pixel size decreases.

Plastics with aromatic monomers like vinyltoluene or styrene such as Saint Gobain BC-420 can produce light output (LO) up to 64 \% of anthracene \cite{wieczorek2017}. Light output and pulse shape discrimination (PSD), which provides separation of gamma ray and neutron interactions, are key metrics for plastic scintillators and are of high importance for neutron imaging. PSD plastics are enabled by high concentrations of primary fluors which generates distinct pulse profiles based on the type of incident radiation \cite{Zaitseva2012}. Common primary fluors or dyes include polyphenyl hydrocarbons, oxazoles, and oxadiazole aryls. Among these, 2,5-diphenyloxazole (PPO) is one of the most widely used due to its superior ability to differentiate between gamma and neutron radiation interactions through PSD \cite{Zaitseva2012}. 

Most traditionally manufactured plastic scintillators are made from aromatic monomers thermally polymerized into large ingots which are then machined down to a chosen geometry. This methodology is generally the most economically feasible method for producing standard plastic scintillator geometries (such as cylinders) and is the lead manufacturing method used by the commercial vendors that produce plastic scintillators \cite{Eljen}. Previous work on fast neutron imagers used thermally cured plastic scintillators for the arrays. For example, one effort constructed an 8 $\times$ 8 array of 13.5 $\times$ 13.5 $\times$ 50 $\text{mm}^3$ optically isolated pixels of EJ-299-34 plastic scintillator \cite{Newby} through cutting, lapping, bonding, and polishing slabs of plastic. Other recent work produced arrays consisting of nine 14.7 $\times$ 14.7 $\times$ 100 $\text{mm}^3$ EJ-276 plastic scintillators coupled with corresponding 16 $\times$ 16 silicon photomultiplier units, arranged in a 3 $\times$ 3 grid \cite{SHUAI2024}. Alternative materials such as organic glass can be either directly cast or subtractively manufactured into 8 $\times$ 8 arrays segmented arrays composed of 6 $\times$ 6 $\times$ 30 $\text{mm}^3$ pixels separated by 1 mm of dead space \cite{Balajthy_2025}. 

In general, neutron imaging resolution is limited by pixel pitch. While finer resolution is theoretically desirable, it is practically constrained by the labor-intensive costly subtractive manufacturing methods required to fabricate such arrays. Although the range of recoil protons in the scintillator pixels suggests a limit on the size of each pixel, it is still possible to decrease the pixel pitch from state-of-the-art arrays before the recoil range becomes an issue. Careful analysis could potentially enable full event reconstruction even in the case that a recoil proton escapes into an adjacent pixel, which could allow for imaging improvements with even smaller pixels. 

However, further reductions in pixel pitch using traditional subtractive fabrication become increasingly impractical due to cumulative machining tolerances, polishing losses, adhesive bond-line thickness, and the introduction of non-scintillating dead volumes between pixels. While use of scintillating fibers is possible, their use in fast neutron imaging is limited by reduced light output per interaction, challenges in achieving uniform optical isolation across large-area arrays, and difficulty in integrating reflector materials without introducing air gaps or optical cross-talk. In contrast, bulk scintillator pixels fabricated as monolithic, optically isolated volumes preserve light yield and enable deterministic control of pixel geometry and inter-pixel boundaries.

Photopolymerization of plastic scintillators allows for significantly faster curing than thermal polymerization, therefore naturally suggesting additive manufacture (AM) of scintillators \cite{Mishnayot2014, Kim2022, Lim2019, Frandsen2023, Dolezal2023, Chandler2022, Chandler2023, Kim2023, Anand2025, MooreCureDye}. In our previous work \cite{Frandsen2023, Dolezal2023, Moore}, we found that through the use of nonaromatic methacrylate and acrylate monomers, solidification time was dramatically reduced to tens of seconds. These nonaromatic resins facilitate high-resolution vat polymerization using commercially available printers \cite{MooreCureDye}. However, vat polymerization printers generally use only one material which complicates the task of printing inherently multi-material (scintillator and reflector) pixel arrays. An additional advantage of vat polymerization is the ability to rapidly modify scintillator formulations by adjusting monomer composition, fluor concentration, or PSD-enhancing additives without full retooling or filament manufacture, which is required for fused filament fabrication (FFF). Unlike FFF-based approaches, vat polymerization does not suffer from inherent inter-filament voids, enabling the production of more optically homogeneous scintillator volumes with minimal internal scattering. These attributes are particularly advantageous for pixelated neutron imagers, where uniform light collection and reproducible PSD performance are critical.

In this work, we present a custom-designed AM assembly to produce high-resolution multi-material pixel arrays as close to fully autonomously as possible. These arrays are manufactured from PSD-capable and photocurable resin which serves as both the active scintillating medium and the adhesive between the plastic scintillator and the reflector, minimizing any dead volume. Alternating print stations cure 3 mm layers of scintillator and bond reflector materials, which results in a growing ``1D'' array of scintillator and reflector layers. These 1D arrays can then be processed into ``2D'' pixelated arrays. The proposed process integrates material deposition and bonding during fabrication unlike the subtractive workflow, where scintillator cutting, reflector placement, adhesive application, and alignment are performed as discrete manual steps. This enables repeatable pixel-to-pixel alignment and more consistent reflector bonding thickness. The additive assembly involves multiple stages designed for automation and scalability, reducing operator-dependent variability and enabling array geometries that would be difficult or cost-prohibitive to realize through conventional machining. 

Details of manufacturing and material properties are presented in this work for both the 1D and 2D arrays. The produced 2D arrays have been characterized for relative light output and PSD capability on a gain-matched pixel-by-pixel basis. 



\section{Materials \& Methods}


\subsection{Scintillator Formulation}
\label{sec:prep}

Components from Sigma-Aldrich to make array formulations include technical-grade isobornyl acrylate (IBOA) containing 200 ppm monomethyl ether hydroquinone as inhibitor, $\geq$ 90\% purity 1,6-hexanediol dimethacrylate (HDDMA) with 75.0 ppm hydroquinone as inhibitor, 99\% purity PPO, and 97\% purity diphenyl(2,4,6-trimethylbenzoyl)phosphine oxide (TPO). BR-541MB difunctional aliphatic polyether urethane methacrylate oligomer (BR-541MB) was sourced from Bomar. 7,7’-Di(4-anisyl)-9,9,9’,9’-tetrapropyl-2,2’-bi-9H-fluorene (Exalite 416) was acquired from Luxottica-Exciton. All chemicals were used as received.

\begin{table}[ht!]
\centering
\caption{Scintillator resin formulation by component fraction}
\label{formula}
\resizebox{0.7\columnwidth}{!}{%
\begin{tabular}{@{}cccccc@{}}
\toprule
\multicolumn{6}{c}{Component Fraction in Formula (wt.\%)} \\ \midrule
BR-541MB & IBOA & HDDMA & PPO & Exalite 416 & TPO \\ \midrule
23.8 & 38.9 & 16.7 & 20 & 0.5 & 0.1 \\ 
\bottomrule
\end{tabular}%
}
\end{table}

The formula used in this investigation is shown in terms of component wt.\% in \Cref{formula} and is a replica of the ``two-dye'' formula described in \cite{MooreCureDye}. This formula has a high loading fraction of primary fluorescent dye PPO to offset the absence of a scintillating aromatic polymer base and ensure PSD \cite{Zaitseva2012}. The secondary dye Exalite 416 was chosen to minimize competition with the photoinitiator for absorbing the 405 nm cure light \cite{Frandsen2023}. The photoinitiator TPO, at a lower concentration of 0.1 wt.\%, initiates polymerization when exposed to light, typically in the ultraviolet (UV) or near-UV spectrum. The remainder of the components (BR-541MB, IBOA, and HDDMA) make up the bulk of the formulation and were optimized for the physical material properties of the solid plastics. When cured in cylindrical test samples (along the lines of the procedure from \cite{Frandsen2023}) this formula yielded solid scintillators with peak spectral responses in the range of 420-430 nm, aligning closely with the ideal wavelength for conventional PMT quantum efficiency. The resulting plastics reach a hardness between 73 and 78 HD, which is compatible with standard sectioning procedures.

The preparation of the resin follows the procedure outlined in previous work \cite{Frandsen2023, MooreCureDye} with minor modifications. Liquid resin batches were prepared by adding solid components (PPO and Exalite 416) to a 300 mL amber glass container, selected to minimize ambient light exposure and prevent premature polymerization. Then, liquid HDDMA and IBOA were added in a 30:70 ratio by weight. The solutions were sonicated at 50$^\circ$C for 30 minutes to facilitate dissolution of the solid reagents. Heating was provided by the sonicator rather than a hot plate as was used in previous work \cite{Frandsen2023,Dolezal2023}. Viscous BR-541 was heated at 100$^\circ$C for 10 minutes to reduce its viscosity and then added to the amber glass container containing the scintillator solution. The solution was then sonicated for another 30 minutes at 50$^\circ$C to ensure thorough mixing. Afterward, the solution was air-cooled and 0.1 wt.\% TPO was added. A final sonication step of 30 minutes was performed in the amber glass container before placing the container and resin under vacuum at 7 mbar overnight to remove dissolved gases. After degassing, the resin was gently poured into a vat within the final manufacturing assembly. Any minor bubble formation or agitation during transfer was allowed to dissipate as the resin rested prior to use.

\subsection{1D Array Automated Assembly}
\label{1DAAM}

\begin{figure}[ht]
	\center
	\includegraphics[width=1\linewidth]{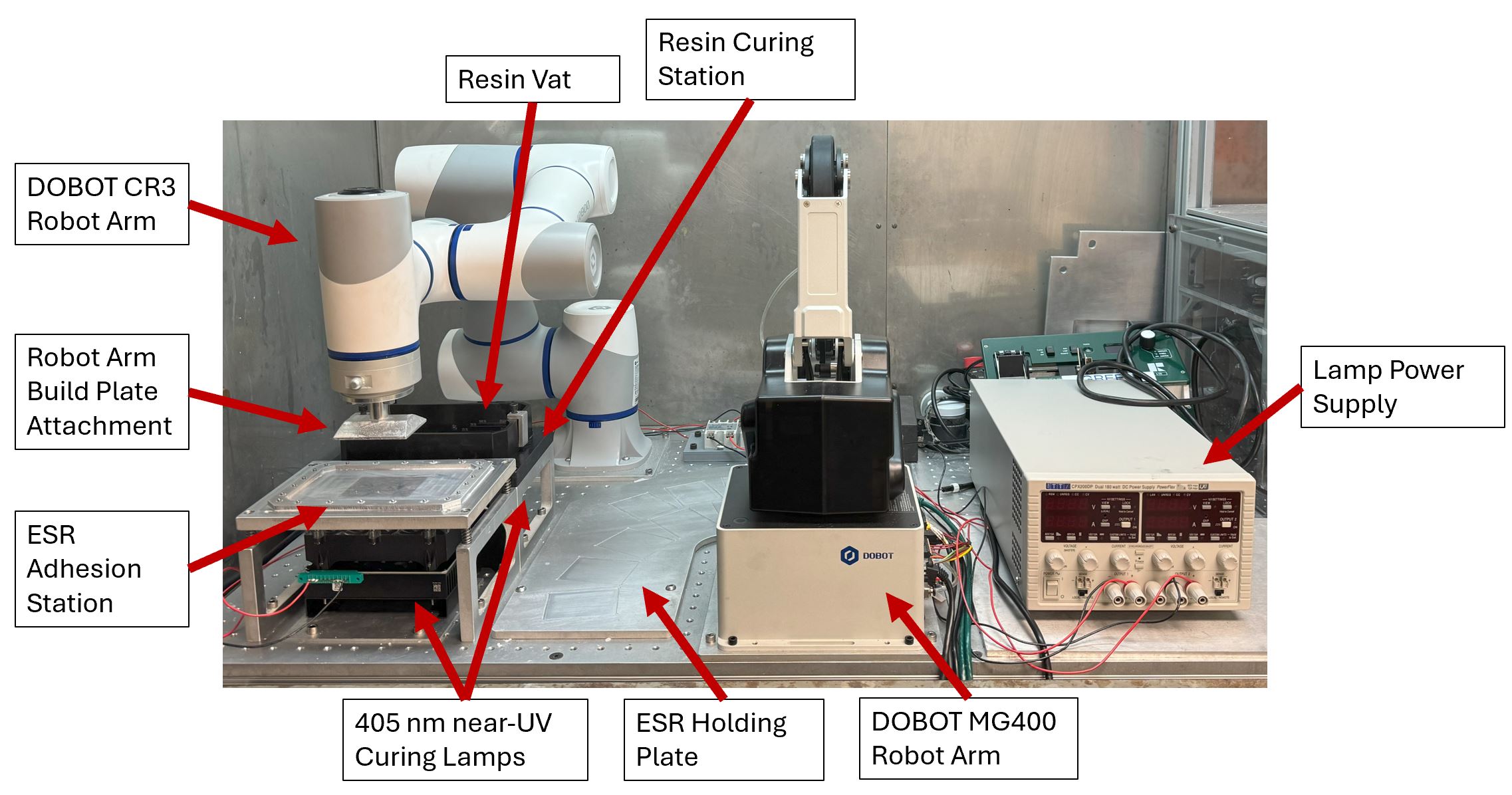}
	\caption{Labeled view of the automated assembly machine. The primary curing station and DOBOT CR3 robot arm are shown in the back left. The DOBOT MG400 robot arm and secondary reflector adhesion station are at forefront center and left.}
    \label{fig:setup}
\end{figure}

The first step in the semi-automated manufacture of 2D plastic scintillator pixel arrays is the autonomous manufacture of 1D arrays. The complete final design is shown in \Cref{fig:setup} and explained below. Various design iterations from earlier stages of development can be found in \cite{Moore}. 

The 1D array automated assembly setup is shown in \Cref{fig:setup}. The setup is based around two robotic arms: a 6-axis DOBOT CR3 and a 4-axis DOBOT MG400. These arms are powered and interconnected through their IO interface for programming and activation. A Prusa resin vat and 405 nm LED assembly were modified for use with an Aim-TTi CPX200DP power supply. The remainder of the brace and housing of the automated arm assembly were modified in collaboration with the Air Force Institute of Technology machine shop. 

\begin{figure}[ht]
    \center
    \includegraphics[width=0.95\linewidth]{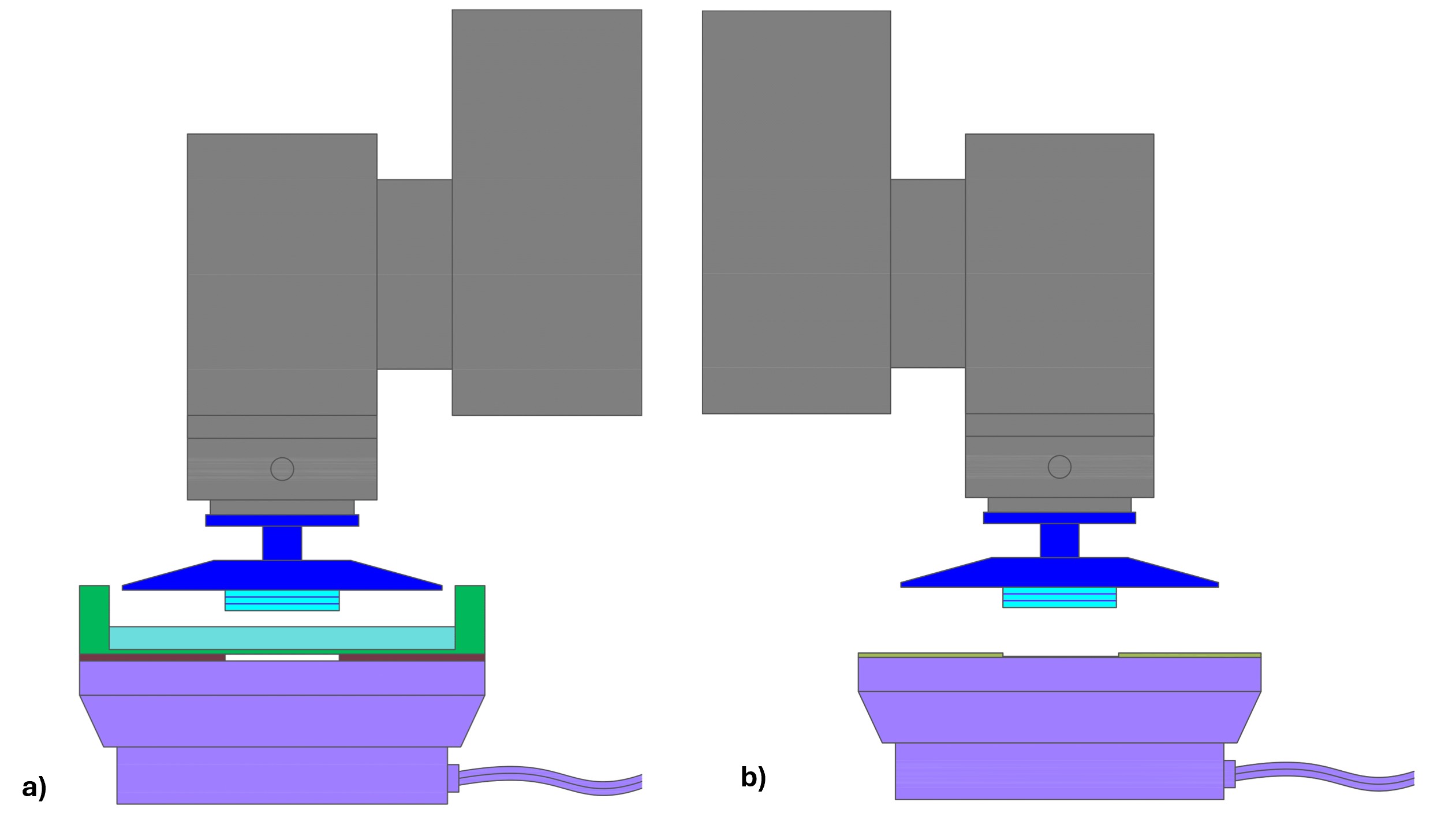}
    \caption{ Automated assembly machine schematic for a) primary curing station and b) reflector curing station. The robot arm and connection point are shown in gray, fabricated build plate in dark blue, resin slabs layers with reflectors in cyan, liquid resin in dark cyan, resin vat in green, light blocker in brown, and 405 nm curing lamp in purple.}
    \label{fig:Setup Schematic}
\end{figure}

In this setup, the DOBOT CR3 robot arm has an attached custom-designed build platform as shown in \Cref{fig:Setup Schematic}a. This platform is lowered into a vat of formula resin with a 55 mm $\times$ 70 mm rectangular hole in the bottom covered by a transparent fluorinated ethylene propylene (FEP) film. The platform enters the liquid slowly and at an angle in order to minimize the production of bubbles in the resin. The vat sits above a Prusa SL1S 405 nm lamp assembly. The platform is positioned approximately 3 mm above the FEP film. This distance is well below the maximum cure depth of the resin under the applied exposure conditions, ensuring complete polymerization through the full thickness of each layer. Then, the lamp is turned on for 60 seconds, triggering photopolymerization in the resin which produces a cured 55 $\times$ 70 $\times$ 3 $\text{mm}^3$ slab of solid plastic. The arm transfers the cured slab to a secondary curing station that features a similar 405 nm lamp. The arm presses the slab upon a piece of 3M Enhanced Specular Reflector Film (ESR) \cite{ESR} that sits on top of the secondary curing station lamp, shown in \Cref{fig:Setup Schematic}b. The lamp brace system utilizes spring elements promoting uniform contact across the interface. The 65 $\mu$m thick ESR reflector foil serves as the optical barrier between each slab of cured scintillator resin. The foil is placed at the secondary station beforehand using the MG400 robot arm and a custom-designed suction diffuser plate to transfer the ESR from a storage area to the cure station without causing deformation. The secondary reflector station lamp then turns on, causing the foil and slab to be bonded and cured together via the polymerization of the residual liquid resin on the slab. The illumination is not directed at a specific incidence angle; rather, the lamp provides upward, diffuse exposure analogous to that used in standard LCD-based vat polymerization systems. The curing light travels through the ESR film, which is highly but not perfectly reflective. Because only a small fraction of the 405 nm light transmits through the ESR, the lamp must stay on for 15 minutes to ensure adequate adhesion. The arm then returns this slab-and-foil layered set to the primary station and positions it approximately 3 mm above the FEP film and a second resin slab cures upon the foil. To mitigate shrinkage or folding of the ESR film, rest periods were incorporated between curing steps to allow stress relaxation prior to subsequent layer formation. The process then repeats until either the resin runs out or the desired number of layers in the 1D array has been completed. 

While primary curing is completed within 60 seconds, transport between stations and static periods to allow samples to cool and resin to rest take an additional 4 minutes per layer. This means for a typical 1D array, the total fabrication time was on the order of 20 minutes per layer set or 220 minutes for 11 layers, dominated by the secondary bonding exposure time. Beyond sample curing, no appreciable increase in resin viscosity was observed during fabrication. Stray polymerization in the vat was minimized by limiting off-axis UV exposure during secondary curing and by returning the layered assembly to the primary station only after lamp shutdown.

The operation of the arms, lamps, and suction diffuser plate is programmed through a Lua script \cite{luaLuaAbout}, a DOBOT-proprietary graphical interface, and manually recorded movement of the arms in DOBOT Studio Pro software. Through this programmed automation, placement errors of the scintillator slabs or ESR layers were not found to occur due to mechanical fixturing and repeatable robot positioning. No manual intervention was required during the fabrication of the arrays reported in this work. 


The physical characteristics of 1D arrays and individual slabs were evaluated, including qualitative characterization of purple discoloration, leaching of the PPO primary dye, bubbling, and quantitative measurements of curvature and dimensional accuracy, as presented in \Cref{result:1D Array Scintillator Fabrication}. These measurements led to changes in the manufacturing procedure to prevent non-representative features that negatively impacted array construction, final LO, and PSD capability. For example, previous work \cite{Frandsen2023, Moore, MooreCureDye} showed that purpling in the samples (which dissipated over the course of several days) was a sign of significant polymerization. Therefore, if this discoloration did not appear during curing, the samples were not fully cured and required additional 405 nm exposure to burn up remaining photoinitiator that would otherwise absorb scintillation light. Measurements of the array slab curvature were an indication of the slab's mechanical rigidity and its adhesion to the build plate during polymerization. Arrays made with early designs of the assembly were highly prone to this curvature. This caused them to fall apart before the completion of successive layering. These observations indicated that the build plate surface was a critical factor in layering success and led to the modification of the manufacturing procedure by sandblasting the build plate to improve adhesion and reduce curvature.

\subsection{2D Pixel Array Conversion}

\begin{figure*}[ht!]
    \centering    
    \includegraphics[width=1\linewidth]{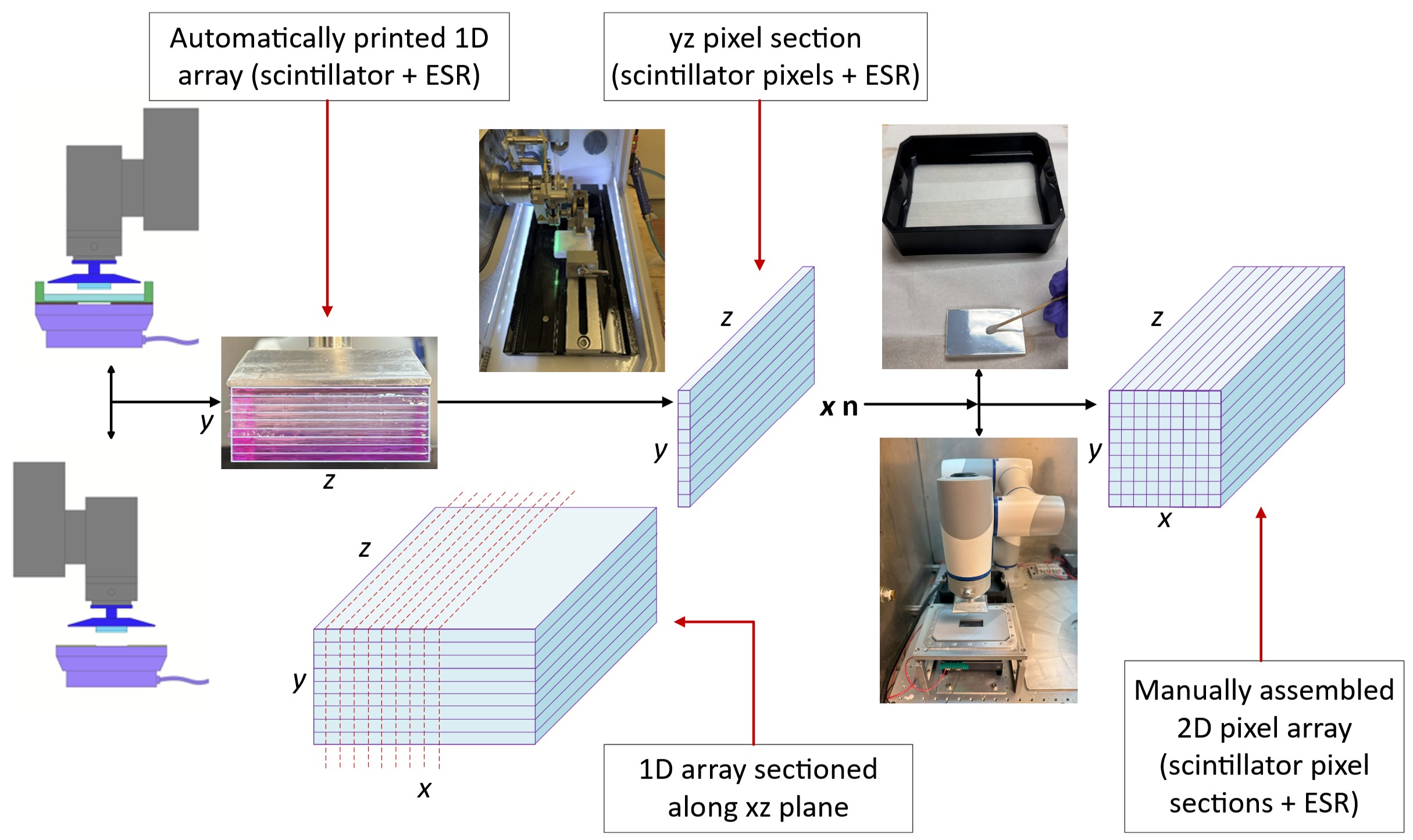}
    \caption{Procedure for construction of 2D-pixel arrays starting with primary curing and ESR application (left), sectioning into pixel layer sections (center), and alignment and pixel array bonding (right).}
    \label{2DArrayConversion}
\end{figure*}

The diagram shown in \Cref{2DArrayConversion} outlines the conversion of completed 1D arrays into 2D pixel arrays. Fully bonded 1D arrays, which already incorporate ESR reflector foil between layers along the stacking vertical $y$ axis, are sectioned along the $xz$ plane using the “repeat cut” function on a Buehler IsoMet High-Speed Pro Automatic Precision Sectioning Machine. The arrays are cut into 3 mm wide sections to match the slab thickness and contain sets of the square pixels shown as the center step in \Cref{2DArrayConversion}. Following sectioning, the pixel sections are inspected for defects, and sections exhibiting imperfections are excluded. ESR reflector foil is applied between the chosen adjacent cut sections along the secondary $yz$ plane face. Uncured scintillator resin from the resin vat serves as the adhesive and is manually applied to these sections. Because the scintillator resin both bonds the sections and constitutes the active scintillating medium, no refractive index discontinuity typical of scintillator–adhesive interfaces is introduced. This results in the only dead volume arising from the ESR reflector thickness. The cut sections are aligned beneath a secondary assembly station and press-bonded using the DOBOT CR3 robotic arm, an alignment fixture, and a 405 nm curing lamp, as shown in the rightmost step of \Cref{2DArrayConversion}. This semi-automated alignment, pressing, and curing process is repeated to assemble the full 2D array. Upon completion, samples are exposed to a final 500 mW/cm$^2$ dose of 405 nm light for 5 minutes in a BlueWave FX-1250 system to polymerize or burn up any remaining TPO along the surface that could reduce light output. During this step, the array is rotated and the exposure side is alternated so that all regions receive an equal dose of UV light, ensuring uniform scintillator curing across the array. Finally, the end faces of the array are cut and polished using a semi-autonomous polishing system to remove surface imperfections down to 1 $\mu$m, ensuring a smooth interface for PMT coupling. All exterior surfaces, except for one end face, are wrapped in ESR reflector foil to enhance light collection efficiency, particularly for outer-layer pixels.

\subsection{Array Characterization}


Each complete 2D array was first evaluated for physical condition, including yellowing/discoloration, transparency, uniformity, and overall dimensional accuracy. Yellowing and general discoloration are an essential concern because they can reduce the LO of individual pixels as well as the array overall. This defect results from the degradation of chromophores in the resin due to heat and oxidation \cite{WU2022, Yousif2013}. Uniformity and overall dimensional accuracy represent the 2D version of the 1D array curvature and dimensional tolerance measurements. This characterization will be impacted by any abnormality or defect in the complete array manufacturing procedure or resin formulation, making it a simple mechanism to demonstrate problems and where changes may be required in the manufacturing methodology. 

\begin{figure}[ht!]
    \centering
    \includegraphics[width=1\linewidth]{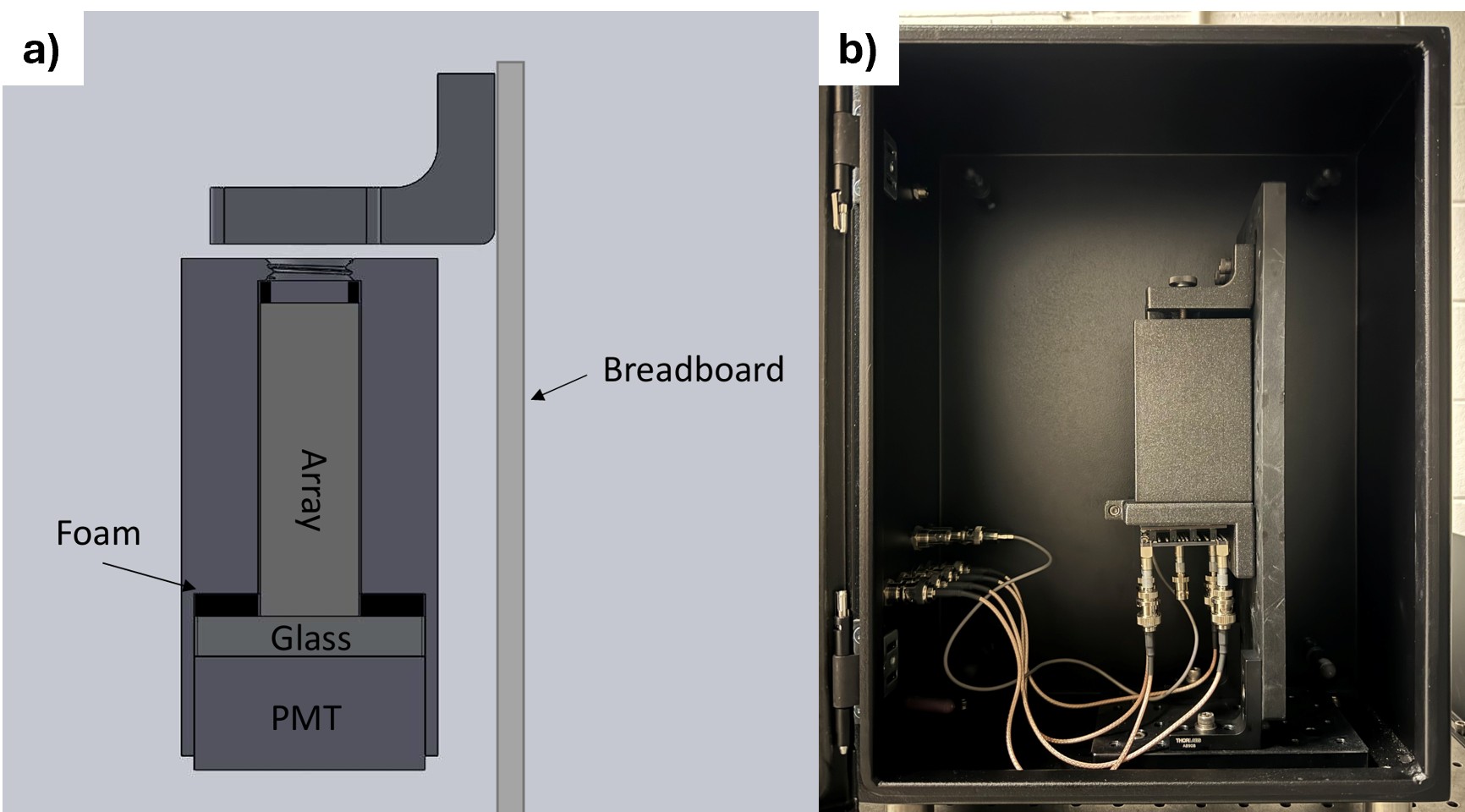}
    \caption{a) Cross-sectional schematic of array holder and b) array holder setup inside a dark box.}
    \label{ArrayBox}
\end{figure}


Array characterization was performed using the experimental design illustrated in \Cref{ArrayBox}a and b. The scintillator arrays were wrapped with seven layers of white Mil-T-27730A polytetrafluoroethylene (PTFE) tape around all sides apart from the one that interfaces with the borosilicate glass window of a Hamamatsu multi-anode H12700 series PMT assembly. While ESR foil provides high reflectivity, the additional PTFE layers further reduce light loss from exterior pixels, improving overall light collection efficiency. This PMT assembly combines 64 anodes into four corner signals through an incorporated signal readout board using resistive division for positional signal processing via the center-of-gravity method. The H12700 multi-anode PMT's 64 anodes are arranged in an 8 $\times$ 8 grid. The inner 6 $\times$ 6 anodes measure 6 $\times$ 6 mm$^2$, while the outer ring of anodes measures 6.25 $\times$ 6.25 mm$^2$. The PMT has a spectral response range of approximately 300–650 nm, which is well matched to the emission spectrum of the plastic scintillator used in this work. A thin layer of Saint-Gobain Crystals BC-630 silicone grease was applied to the PMT window for index of refraction matching. Then, a custom-designed and 3D-printed holder is pressed down around this assembly along the side of the PMT. This creates a light-tight seal and securely holds the scintillator array to the face of the PMT as shown in \Cref{ArrayBox}a. The 3D-printed holder allows the pixel arrays to be utilized inside and outside the dark box while ensuring no background light reaches the PMT. A 5 $\mu$Ci $^{207}$Bi source was aligned with the array and PMT face for measurement at 1'' from the detector. This source has a thin window, allowing for the escape of conversion electrons in addition to the standard gamma rays associated with $^{207}$Bi. A CAEN Technologies DT5730 digitizer recorded the pulse signals produced from each corner output of the PMT assembly. The trigger condition enforced coincidence conditions of signals in each of the four corners occurring within 20 ns of each other. Each corner pulse is integrated over a 300 ns window, and the sum of the four corner integrals provides a measure of the detected light amplitude. An amplitude with a shorter integration time of 30 ns was also calculated for pulse shape discrimination. Modified weighting Anger logic equations from the H12700 manual \cite{hamamatsu} were applied to valid signals to compute the corresponding $X$ and $Y$ pixel positions. These positions are determined from the resistive division of the multi-anode PMT corner signals ($A$, $B$, $C$, and $D$) \cite{Anger1964, Peterson2011}:

\begin{equation}
X = \frac{B + D}{A + B + C + D}, \qquad
Y = \frac{C + D}{A + B + C + D}
\label{XYEQ}
\end{equation}

\noindent Using the reconstructed $X$ and $Y$ coordinates, events from the same acquisition and detector–source geometry were segmented into individual pixels by applying fixed spatial cuts corresponding to the predefined pixel boundaries. For each event assigned to a given pixel, the pulse integral spectrum was constructed from the total integrated signal, defined as the sum of the four corner integrals

\begin{equation}
E_{\text{total}} = A + B + C + D
\label{Etotal}
\end{equation}
\noindent which provides a measure of the detected scintillation light.

Despite having equivalent external dimensions, individual pixels can exhibit different light output and detection efficiency. These variations can arise from several factors, including position-dependent light collection efficiency, particularly for edge and corner pixels, due to optical transport effects and reflector properties \cite{Newby}. For arrays where each pixel has an independent photosensor with adjustable bias voltage, discrepancies can be corrected via hardware gain adjustments. In the present case, where multiple small pixels are coupled to a single multi-anode PMT with a common bias, hardware adjustments are not possible. Instead, each pixel’s response was individually gain-matched in software. The signal from each pixel under irradiation by a known radiation source was compared to a reference pixel, and a correction factor was determined by taking the ratio of each pixel's channel number for characteristic spectral features such as Compton edges and electron peaks. This factor was then applied to scale the pixel signals, effectively normalizing all pixels to a common light output level, ensuring consistent response across the array. All analysis was performed using C++ code with the ROOT framework \cite{root}.



The same setup was used for characterization of the array's radiation response to an Americium-Beryllium (AmBe) source, which supplies both neutrons and gammas. Sample responses were recorded using a CAEN DT5730 digitizer with a CAEN DT8033M HV power supply module at -1,050 V bias voltage. Recording and control were performed using CoMPASS data acquisition software and GECO voltage control software, respectively. The bias voltage was chosen to maximize signal gain without exceeding the rated limits of the multi-anode PMT \cite{hamamatsu}.

For each event, the PSD parameter was calculated using the standard definition,
\begin{equation}
\label{PSD}
\text{PSD} = \frac{E_\text{long} - E_\text{short}}{E_\text{long}},
\end{equation}
where $E_\text{long}$ and $E_\text{short}$ are obtained from the sum of the four corner signals. For PSD analysis of individual pixels, events were assigned to pixels based on fixed spatial cuts corresponding to the predefined pixel boundaries. 

The overall PSD capability of each pixel was then assessed by calculating the Figure-of-Merit (FoM), determined from the centroids of the gamma and neutron PSD distributions ($\mu_n$, $\mu_{\gamma}$) along with their respective Full Width at Half Maximum (FWHM) ($FWHM_n$, $FWHM_{\gamma}$):
\begin{equation}
\label{FOM}
\text{FoM} = \frac{\mu_\text{n} - \mu_{\gamma}}{\text{FWHM}_\text{n} + \text{FWHM}_{\gamma}}.
\end{equation}
\noindent where the centroids $\mu_n$ and $\mu_{\gamma}$ and their corresponding FWHMs were obtained by fitting Gaussian functions to the neutron and gamma peaks in the one-dimensional PSD histograms for each pixel.

\section{Results and Discussion}
\label{sec:results}

\subsection{1D Array Scintillator Fabrication}
\label{result:1D Array Scintillator Fabrication}

\begin{figure}[ht!]
    \centering
    \includegraphics[width=1\linewidth]{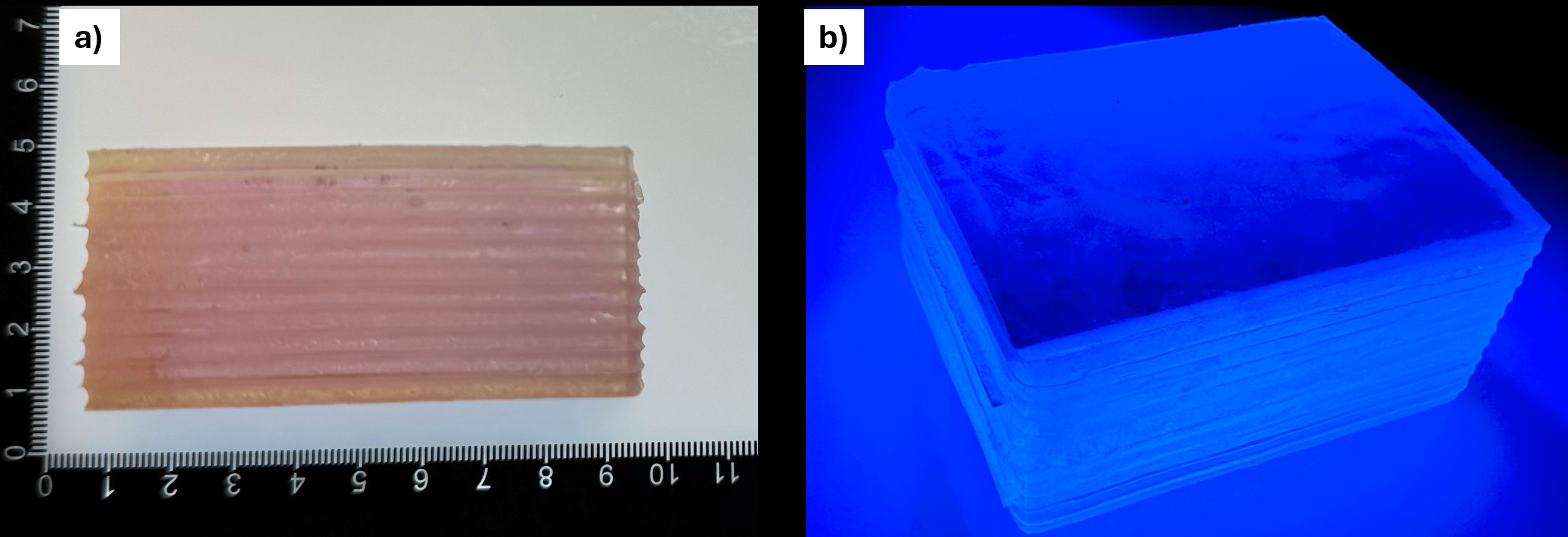}
    \caption{11-layer 1D scintillator array a) side profile immediately after curing and b) under 405 nm light.}
    \label{1DArray}
\end{figure}

Following the outlined autonomous manufacturing process described in \Cref{1DAAM}, 1D scintillator arrays of up to eleven layers (each 3 mm thick) were made using a single 200 g batch of scintillator resin. This maximum size of the array (shown under normal and 405 nm light in \Cref{1DArray}) is currently limited by the maximum capacity of the curing station resin vat. The uppermost and lowermost layers contain the highest concentration of deformities, including surface roughness and bubbles, respectively. For this reason, these layers are either stripped off in the case of the first layer or simply not produced in the case of layer eleven. The surface roughness observed in the first layer is limited to the surface bonded to the build plate. This roughness results from the sandblasting of the build plate, which is required for sufficient adhesion to support the array and prevent peeling. Bubbling and gaps are most common in the final layer due to insufficient resin available in the vat and repeated agitation and introduction of air into the resin during movement into and out of the vat. This limitation means the largest usable array possible with the current resin vat and 200 g resin batch setup is nine 3-mm-thick layers of scintillator with a layer of ESR between each.

In \Cref{1DArray}a and b, a small amount of excess polymerization is observed along the edges of the slabs. This excess polymerization represents areas where the resin solidified and cured beyond the designed 55 $\times$ 70 $\text{mm}^2$ area. One approach to dealing with this excess polymerization is to use a tertiary cure-depth-limiting dye in the resin formulation to shorten the penetration depth of the 405 nm cure light \cite{MooreCureDye}. However, restricting the cure depth would prevent complete polymerization of the 3 mm-thick slabs and compromise bonding to ESR layers. Instead, excess polymer is removed during the conversion of 1D arrays into 2D pixel arrays.

\subsubsection{Purple Discoloration}

\begin{figure}[ht!]
    \centering
    \includegraphics[width=1\linewidth]{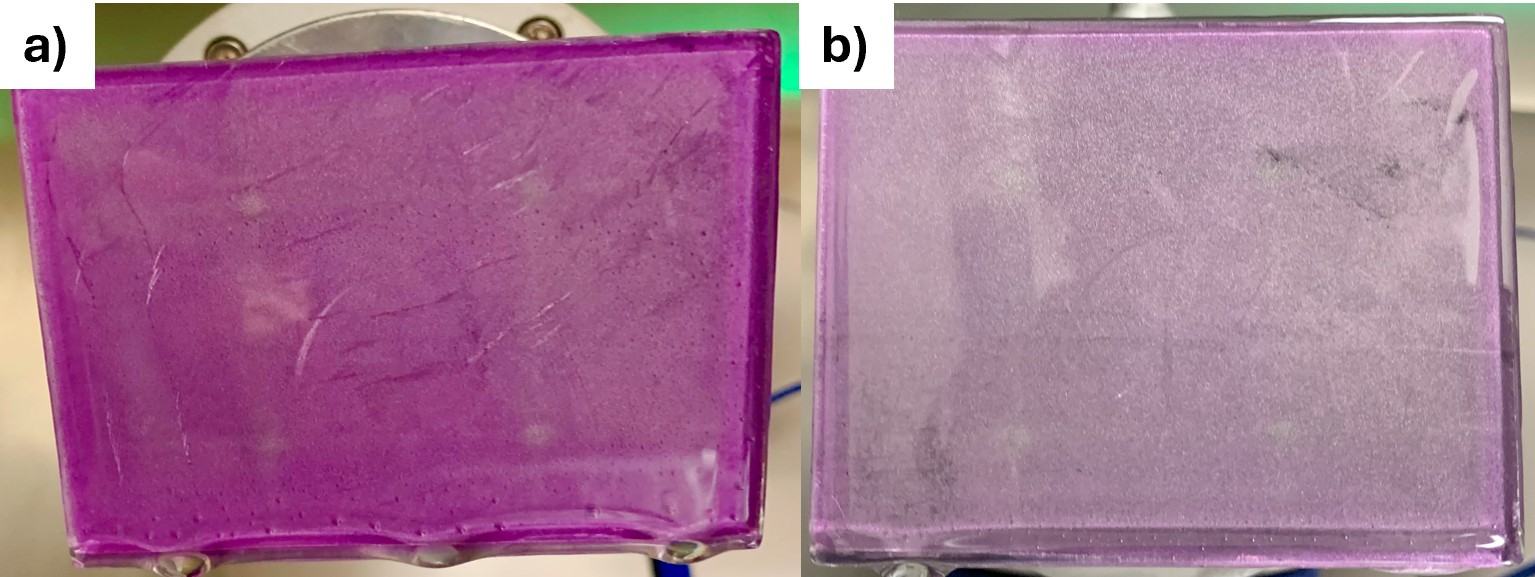}
    \caption{Cured 3 mm scintillator resin slabs. The purple color fades over time, with high intensity a) immediately after curing and low intensity b) 3 hours post-curing.}
    \label{Purple}
\end{figure}
 
A purple discoloration was observed in all array layers during and after the photopolymerization process, as shown in \Cref{Purple}. This discoloration occurred as quickly as after 30 seconds of exposure to a fluence of 1.3 J/cm$^2$ of curing and became more intense as the curing progressed. A small amount of patterning from the lamp light segmentation is apparent in \Cref{Purple}a but this quickly dissipates. The purple color intensity decreases over time, with purple color significantly reduced over 3 hours, as shown in \Cref{Purple}b, and entirely dissipated over 48 hours. 

The appearance of purpling is likely a result of the polymerization interaction between PPO and TPO when exposed to UV and near-UV light \cite{Frandsen2023}, making it visible only in areas where substantial curing has taken place. Prior research has also demonstrated that the fading of the purple hue can be accelerated using an oven and elevated temperatures \cite{Frandsen2023}. However, this study did not investigate heating to avoid additional heat-induced discoloration.

\subsubsection{Leaching}

\begin{figure}[ht!]
    \centering
    \includegraphics[width=1\linewidth]{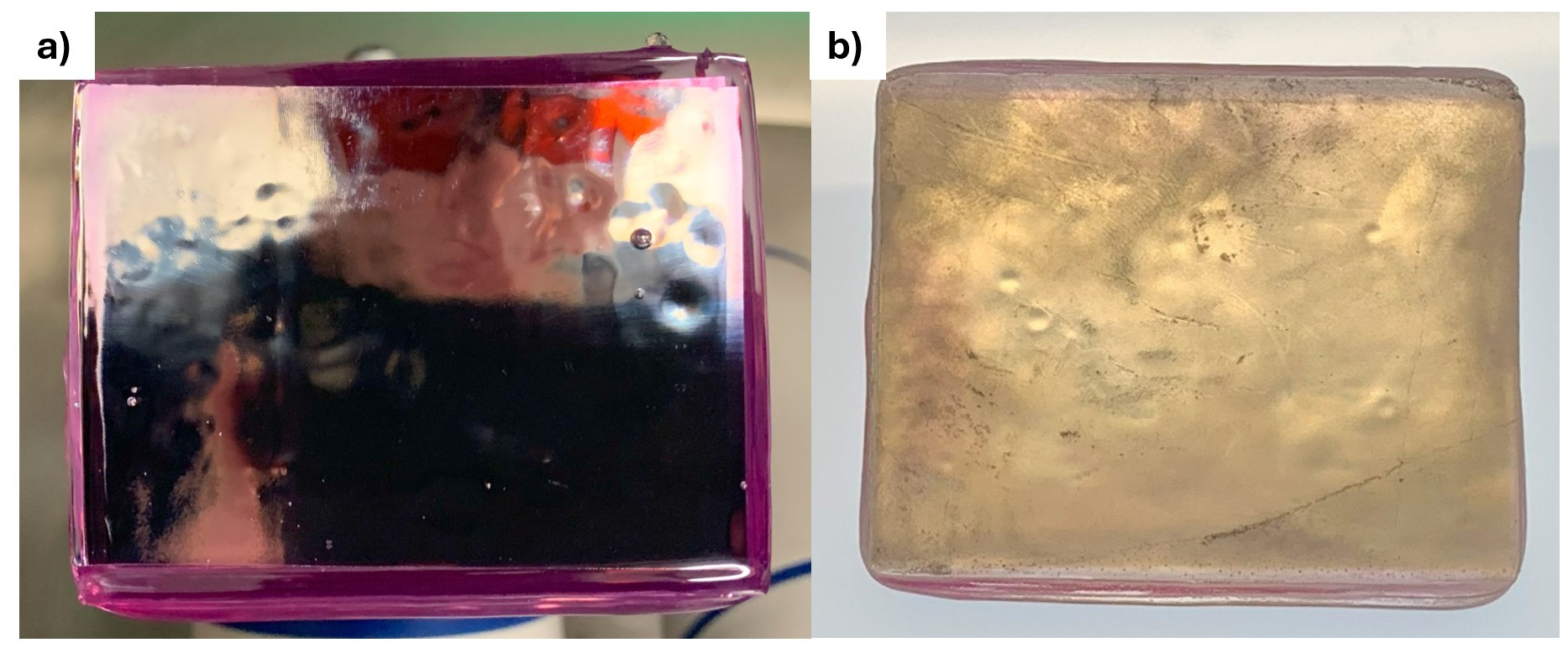}
    \caption{Cured 1D array leaching over time. The haze intensity increased with time from a) no visible fogging immediately after curing to b) complete fogging by 36 hours post-curing.}
    \label{Leach}
\end{figure}

Leaching of fluorescent compounds was observed to occur along the outer surfaces of all sample arrays over 48 hours. Previous studies on cylindrical photocured scintillators have indicated that leaching is generally not observed in formulations containing 15 wt.\% or less PPO \cite{Frandsen2023, Dolezal2023}. Other work has demonstrated that samples with PPO loadings of 20 wt.\% or higher consistently exhibit leaching, leading to the formation of a hazy, white film on exposed scintillator surfaces \cite{Frandsen2023, MooreCureDye, Chandler2023}. \Cref{Leach} shows that in this case leaching was evident along flat exposed surfaces. A slight haze appears after 24 hours and substantial surface fogging develops after 36 hours, as depicted in \Cref{Leach}b. In contrast, no leaching or haziness was observed at the ESR-resin slab interfaces internal to the array.

The observed leaching results from PPO incorporated into the resin at concentrations near its solubility limit. As PPO can migrate through the polymer matrix upon solidification, a concentration gradient can form at surfaces exposed to oxygen. This drives diffusion of PPO toward the exterior surface of the array. Upon reaching the surface, interaction with oxygen, moisture, or ambient light can further accelerate migration and promote crystallization, appearing visually as a hazy or cloudy layer \cite{Broughton2023}. In contrast, internal regions of the array, such as the ESR-resin interfaces, remain effectively sealed from the external environment. In these areas, PPO mobility is significantly restricted due to the absence of a concentration gradient and the physical encapsulation provided by surrounding materials. Additionally, the lack of exposure to atmospheric oxygen, moisture, and light prevents the initiation of leaching or crystallization processes, ensuring that no visible haziness develops internally.

Cleaning the arrays with 91\% isopropyl alcohol, deionized water, or a combination of both restored the original transparent appearance of the arrays. However, the hazy film reappeared once the resin slabs dried either by hand or air, typically within 24 hours. Previous work \cite{Frandsen2023, Zaitseva2018} mitigated leaching through prolonged immersion (greater than 1 hour) in 70\% or greater ethanol baths. This method has been shown to produce minimal loss in material and delay the return of hazy discoloration for up to two months. Ethanol immersion was not incorporated in this work due to uncertainty regarding its impact on scintillator-reflector bonds. Instead, arrays were cleaned immediately prior to characterization and measured before the reappearance of the hazy surface film. Additionally, repeated cleaning cycles were observed to reduce the severity of the haze upon subsequent drying.

\subsubsection{Dimensional Uniformity}

\begin{figure}[ht!] 
\centering
\includegraphics[width = 0.5\columnwidth]{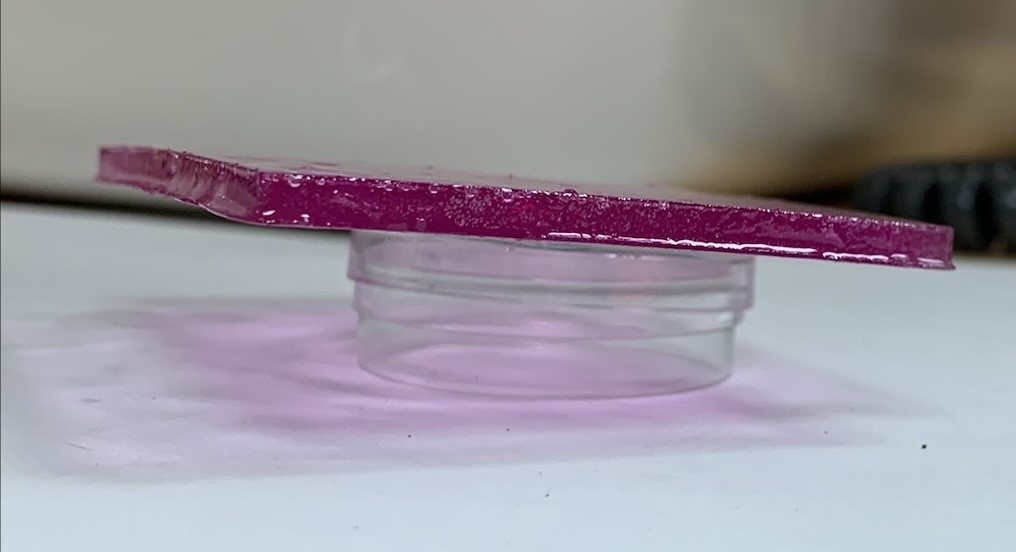}
\caption{Single scintillator slab viewed from the side. Warping is visible along both the width and length axis.}
\label{SlabCurve}
\end{figure}

Five-layer arrays were measured to gauge the consistency and accuracy of the physical dimensions of interest. The design dimensions of each resin slab were 55 $\times$ 70 $\text{mm}^2$ in area (set by the build plate attachment shape) and 3 mm in thickness (set by the distance between the build plate and the FEP window). Initial designs featured an oversized build plate but resulted in significant excess polymerization beyond the 55 $\times$ 70 $\text{mm}^2$ area. This effect was significantly reduced but not entirely eliminated by reducing the size of the build plate to match the desired area.

\begin{figure}[ht!]
    \centering
    \includegraphics[width=0.7\linewidth]{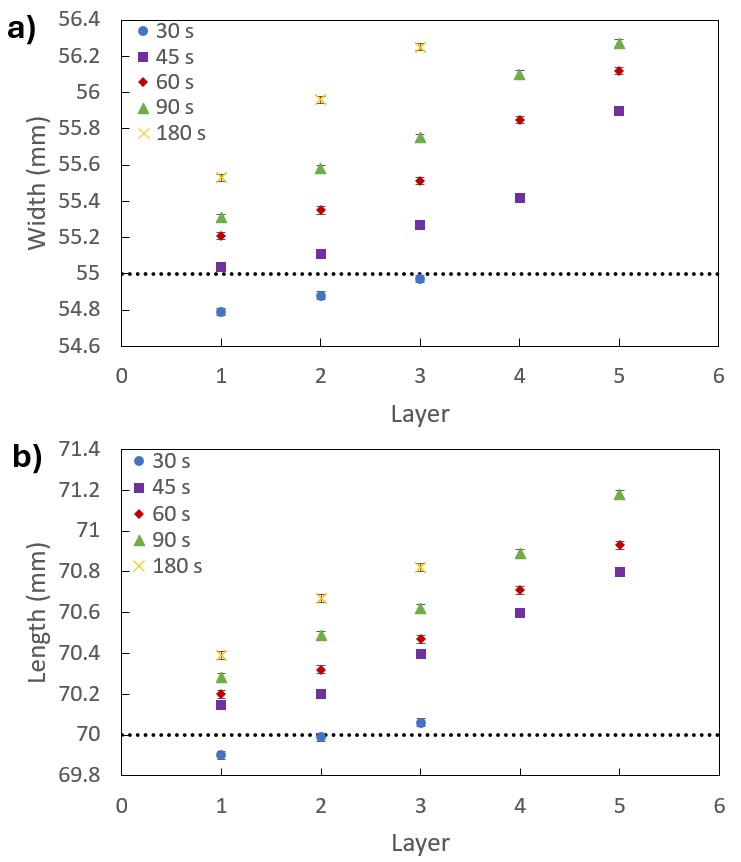}
    \caption{Measured a) width and b) length of cured 3 mm thick slabs for layer curing times of 30, 45, 60, 90, and 180 s/layer. Expected width and length are shown as dashed black lines. Error bars represent the caliper measurement uncertainty ($\pm$0.02 mm) and are smaller than the marker size at this scale.}
    \label{dimension}
\end{figure}
 
The degree of remaining excess polymerization correlated with the per-layer curing time. The exact dimensions of the arrays along the $x$ (width) and $z$ (length) axes were measured for layers 1 to 5 of a 1D array assembly for five different curing times. \Cref{dimension} shows results for the measured curing times 30, 45, 60, 90, and 180 seconds. Each data point represents the dimension of the corresponding individual resin layer for each curing time. The 30- and 180-second curing times end at the third layer. This smaller number of layers results from different mechanisms depending on the curing time. For a 30-second cure, the layers peel apart, whereas for a 180-second cure, excessive polymerization causes the slabs to grow beyond their intended dimensions and become heavy, leading the upper layers to bend and separate. Curing at 45, 60, and 90 s per layer was completed up to or beyond five layers. All curing times show a general upward trend as the layers are built, indicating increasing excess polymerization beyond the desired 55 $\times$ 70 $\text{mm}^2$ shown as the dashed black lines in \Cref{dimension}. The excess polymerization increases with each layer due to the additional region of polymerization from the previous layer providing more area for polymerization to build on. The 45, 60, and 90 s per layer curing times had sufficiently low over-polymerization that repeated layering proved successful. This indicated viability for these values as per-layer curing times. While 45 s per layer initially appears to be the most optimal due to its low amount of excess polymerization, over-polymerization is not the only consideration when choosing a layering time. Repeatability in designed thickness and lack of curvature also play a major role. 

\begin{figure}[ht!]
    \centering
    \includegraphics[width=0.7\linewidth]{Figures/ArrayThick.jpg}
    \caption{Measured layer thicknesses for the 5-layer array using layer curing times of 45, 60, and 90 s/layer. Error bars represent the micrometer measurement uncertainty ($\pm$0.002 mm) and are smaller than the marker size at this scale.}
    \label{thickness}
\end{figure}
 
\Cref{thickness} shows the measured thickness of the first five layers averaged across three positions along the surface of the layer for 45, 60, and 90 seconds per layer curing times. All three formulas are within 0.15 mm of the desired 3 mm thickness. The 45 s curing time results show thicknesses 0.1-0.15 mm below the design value, while the 90 s curing time shows measurements slightly above the desired thickness. The 60 s cure data sit between the two, only slightly below 3 mm on average. One explanation for the scatter in thickness is that there is a loss in the accuracy of the robot arm over repeated movement. This leads to variability in the layer thickness, causing each layer to be improperly sized. Repeatability tests with this specific robot arm in all three dimensions revealed that the discrepancy between the expected and measured distances was an order of magnitude higher for the vertical dimension (0.2 mm) than for the horizontal dimensions (0.02 mm). Curing time also plays a role in how thick the layers grow. One explanation is that at higher curing times, the lamps generate more heat. This heat within the curing layers cause them to expand into the FEP below causing the layer itself to grow thicker. Another explanation is due to the increased heat, the FEP becomes more pliable which allows the resin more room to grow larger. Despite potential differences in UV exposure between early and late layers, the high reflectivity of the ESR foils and the penetration depth of the light into the resin ensure largely uniform polymerization across all layers, supporting consistent layer quality.

\begin{figure}[ht!]
    \centering
    \includegraphics[width=0.7\linewidth]{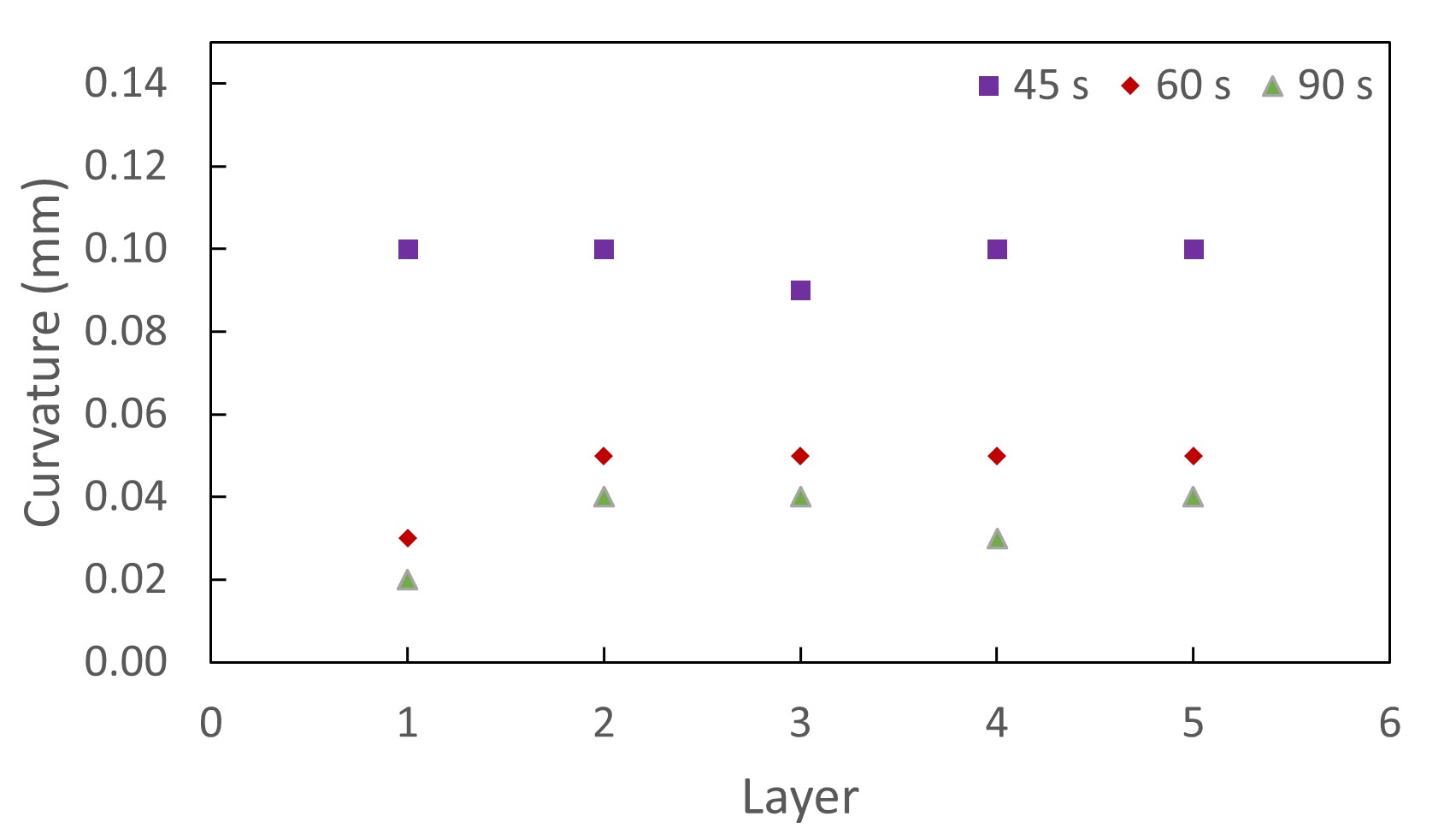}
    \caption{Measured curvature of cured 3 mm resin slabs using layer curing times of 45, 60, and 90 s/layer.}
\label{Curvature}
\end{figure}


The plastic slabs also displayed curvature or warping after curing. \Cref{SlabCurve} illustrates an extreme case of curvature in an array cured for 45 seconds, where the corners bent away from the build plate. This slab curvature was measured by subtracting the thickness at a corner from that at the center. \Cref{Curvature} shows the results of this measurement technique. Per-layer curing times of 60 and 90 seconds have equal to or less than 0.05 mm of curvature, while 45 seconds has around 0.1 mm of curvature. One explanation is that when the build plate pulls up from the FEP, the less well-cured and softer slab corners peel away, and when the slab cools and shrinks the curvature intensifies. This curvature indicates that longer curing times allow for more well-cured slabs and better adhesion and, therefore, more viability for larger, well-defined arrays. A 60 s per-layer exposure was therefore selected as the optimal condition for array fabrication, as it produced a similar degree of curvature while reducing excess polymerization relative to the 90 s per-layer curing time.

\subsubsection{Adhesion Properties}

\begin{figure}[ht!] 
\centering
\includegraphics[width = 0.7\columnwidth]{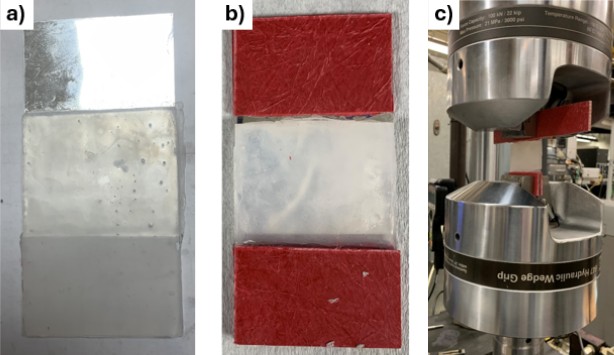}
\caption{Scintillator slab adhered to ESR foil sheet with 20 mm of overlapped bonded material a) before fiberglass attachment, b) with fiberglass attachment, and c) in MTS shear strength testing equipment.}
\label{MTS}
\end{figure}

The resin must be able to create a strong bond to the reflector material when cured in order for the array to maintain structural integrity. Use of the scintillator resin itself as the bonding agent reduces the dead area on the pixel array and avoids issues with a refractive index mismatch that hurts light collection. However, the primary metric for an adhesive is its bonding strength, so there is only value in using the scintillator resin in this way if the bond is sufficiently strong. The bonding performance of the resin was compared to that of a commercial UV-curable adhesive (OP-20 from Dymax) between solid scintillator slabs and the ESR foils. A total of five samples were fabricated and evaluated following ASTM Standard D1002 Shear Testing \cite{ASTMD1002} utilizing the MTS Acumen Electrodynamic Test System and shown in \Cref{MTS}. Three samples were prepared and compared in parallel with commercial OP-20 UV-curable adhesive replacing the novel resin. This was done to account for the unknown target bond shear strength value prior to testing. The resin slab and ESR overlap was 20 mm, with 25.2 mm of material glued to fiberglass grips held within the shear strength tester. 



Following adhesion testing, the scintillator resin was found to exhibit shear strength on the same order of magnitude as OP-20, though with somewhat lower peak values. Resin-bonded samples required an average separation force of approximately 230 N, compared to over 330 N for OP-20. Notably, adhesion strength for the resin–ESR interface decreased with increased time between bonding and testing. Samples bonded approximately two months prior to testing, exhibited lower shear strength than samples tested within one week of bonding. When only resin samples prepared within one week of testing are considered, the average separation force increases to above 290 N for the resin-bonded samples. This trend suggests time-dependent degradation at the resin–ESR interface. Possible mechanisms include incomplete curing allowing polymer relaxation, interfacial stress development due to differential shrinkage, or gradual loss of secondary bonding interactions at the polymer–ESR interface. The data indicates that either extended UV post-curing or surface treatment optimization may be required to maximize long-term bond durability.

Although OP-20 demonstrated higher peak shear strength, the scintillator resin provides sufficient adhesion for array-scale fabrication while offering the additional advantage of functioning as both the active scintillating medium and the bonding agent.

\subsection{2D Pixel Array Scintillator Fabrication}

\begin{figure}[ht!]
    \centering
    \includegraphics[width=0.7\linewidth]{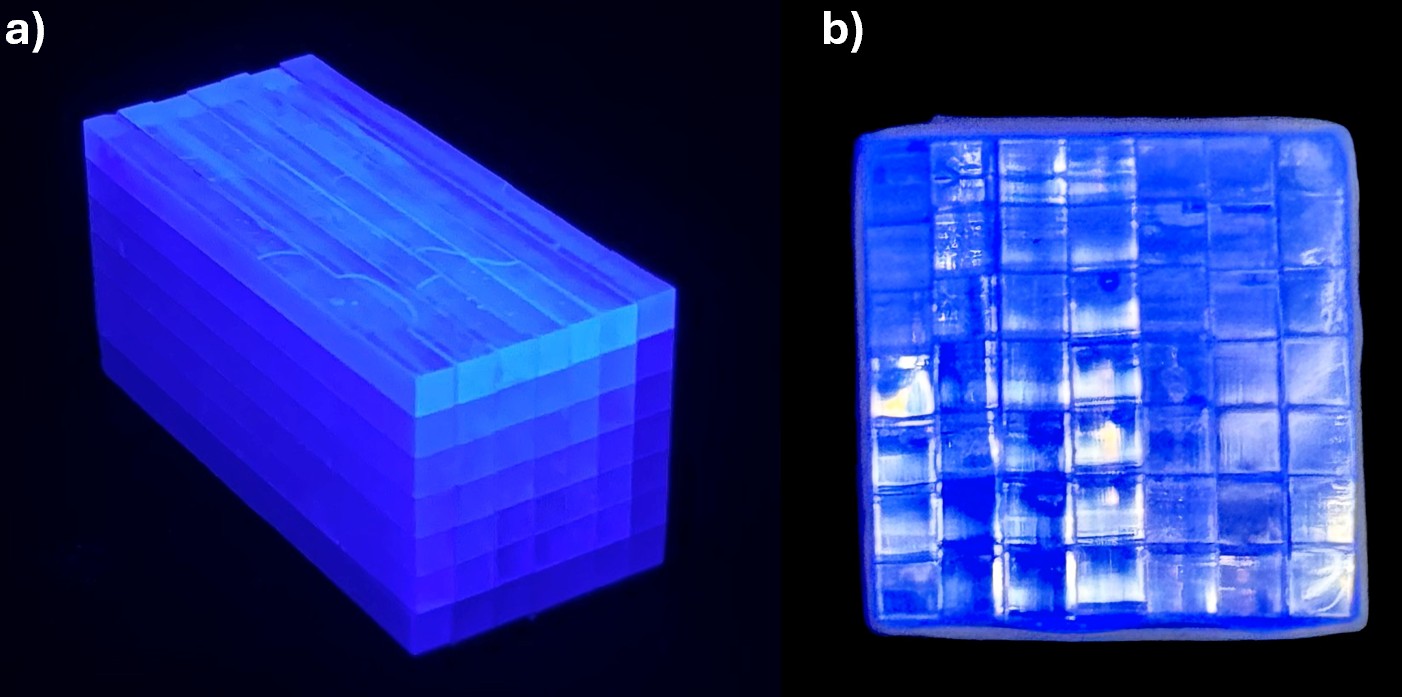}
    \caption{Complete 7 $\times$ 7 2D pixel array a) length view and b) pixel face view under 405 nm illumination.}
    \label{2DArray}
\end{figure}

Following autonomous 1D fabrication, 2D pixel array conversion was performed according to the methodology described in \Cref{2DArrayConversion}. This approach enabled the production of arrays up to 7 $\times$ 7 pixels and 70 mm in length. Although the automated system produced up to nine well formed and bonded layers per 200 g resin batch during 1D fabrication, the final 2D array, shown under 405 nm illumination in \Cref{2DArray}, contains seven layers in both the x- and y-directions. This reduction reflects limitations associated with the manual steps required for 1D-to-2D conversion, including pixel alignment, ESR insertion, bonding, and polishing following 1D array sectioning. Automation significantly reduces handling during array fabrication by bonding ESR directly to freshly printed scintillator layers using residual uncured resin. However, each interface in the 2D assembly must still be processed individually due to the limited penetration depth of 405 nm light and the presence of reflective ESR foils, which preclude uniform post-curing of stacked arrays. As a result, pixel sections exhibiting visible defects were preferentially excluded to maximize array uniformity. For example, one slab at the bottom of the array contained localized bubbling ($\le$1 mm) at the ESR–resin interface, attributed to incomplete wetting during bonding. This defect occurred only when the remaining resin volume was very low, and the affected pixel section was therefore excluded. An additional section was removed due to surface scorching along one edge, likely caused by excessive friction during sectioning as coolant flow became insufficient at the selected feed rate. Although such damage can be remediated through polishing, this step was deemed inefficient given the availability of defect-free pixel sections.

While the initial 2D-pixel array assessed was 70 mm long, this array was later cut into the 50 mm array shown in \Cref{2DArray}a and b and a 20 mm array for further analysis of transparency and position resolution. No delamination or peeling was observed upon cutting the array into smaller sizes. In \Cref{2DArray}b, a small amount of pixel misalignment is observed in the far-left column. This offset results from the manual placement of the pixel slabs during alignment under the curing press, shown in the rightmost step of \Cref{2DArrayConversion}. The effect is evident in the position resolution measurements when the array is irradiated.

\subsubsection{Complete Array Discoloration}

\begin{figure}[ht!]
    \centering
    \includegraphics[width=.7\linewidth]{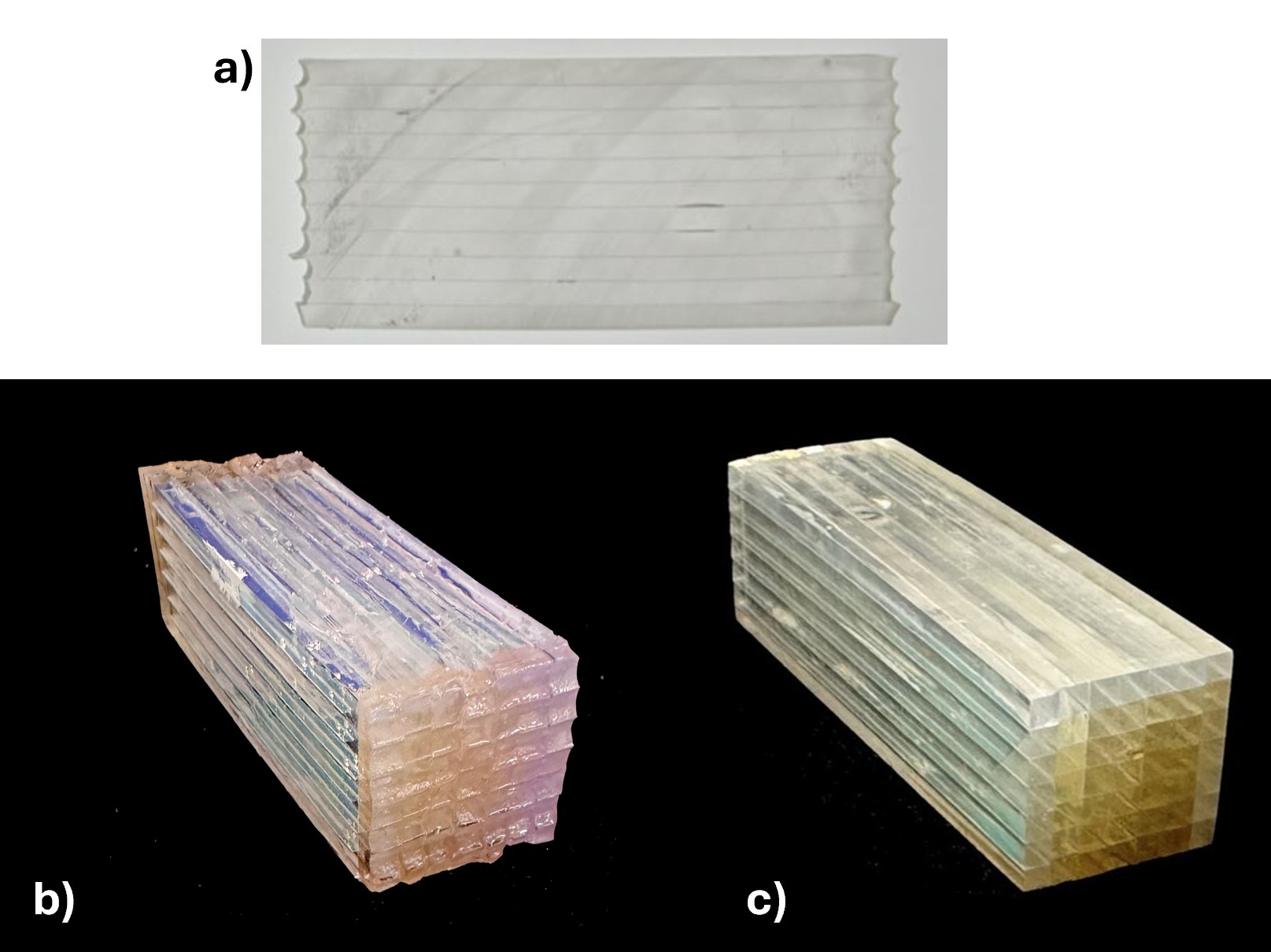}
    \caption{2D Array discoloration a) 48 hours after post-curing of 1D array and sectioning, b) immediately after incremental post-curing of one side of the 2D array, and c) 48 hours after post-curing and polishing of the 2D array.}
    \label{Discoloration}
\end{figure}

As in the 1D arrays, discoloration was present in the final 2D array. Shown in \Cref{Discoloration} is the changing of colors of the array throughout its conversion from a 1D array to the final 2D-pixel array. The array layers were initially a deep purple, which dissipated over approximately 48 hours, producing the colorless pixel slab shown in \Cref{Discoloration}a. After assembling and bond-curing the full array, followed by post-curing with 405 nm light, a weaker purple coloration reappears throughout the array, with \Cref{Discoloration}b capturing an intermediate stage where only one side had been illuminated. This re-emergence of purple color indicates the 1D array layers were not entirely cured by the time of re-bonding and contained additional TPO that required burn up or removal with post-processing. This new purple coloration also dissipates with time, disappearing entirely from the array in under 48 hours, as shown in \Cref{Discoloration}c.

However, unlike within the 1D array manufacturing, the dissipation of purple color leaves minor yellow discoloration within certain pixels as evidenced by the lower center pixels of \Cref{Discoloration}c. While the dark background of the photograph intensifies the discoloration, the yellow color was present in all sample arrays manufactured. This yellow coloration was analyzed in previous work \cite{MooreCureDye} to be the result of excess heat during or post-polymerization. This heat was hypothesized to thermally oxidize the polymers or otherwise induce chromophore formation in the samples, resulting in discoloration. In previous work, this effect was mitigated through cooling baths and modifications to curing parameters; however, reducing post-curing exposure was found to have the most significant impact for these arrays. Yellowing intensified most strongly following exposure in the BlueWave FX-1250, where sample temperatures exceeded 60$^\circ$C. To prevent overheating, pauses were introduced after every minute of exposure to allow the samples to cool to room temperature.

\subsubsection{Transparency}

\begin{figure}[ht!]
    \centering
    \includegraphics[width=0.55\linewidth]{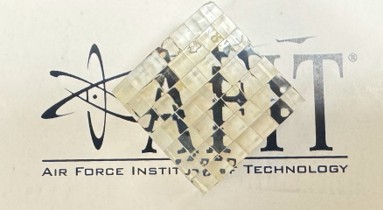}
    \caption{Array pixel face transparency. Light yellow discoloration is seen in several pixels.}
    \label{20mmface}
\end{figure}

\Cref{20mmface} shows the 20 mm array placed over a card with complex lettering. The array is highly transparent and the ``AFIT'' lettering is easy to distinguish. Some pixels appear to slightly distort the image, resulting from internal reflection and the angle at which the photo was taken.

\subsubsection{Dimensional Accuracy}



\begin{table}[ht!]
\caption{Measured 2D array dimensions averaged over five measurements and labeled according to nominal length. The width corresponds to the dimension of the printed 1D array.}
\centering
\label{ArrayDim}
\resizebox{1\columnwidth}{!}{%
\begin{tabular}{lcllcllcll}
\hline
\multicolumn{1}{c}{Array Type} & \multicolumn{3}{c}{Width (mm)}         & \multicolumn{3}{c}{Height (mm)}         & \multicolumn{3}{c}{Length (mm)}         \\
\multicolumn{1}{c}{}           & Projected & \multicolumn{2}{c}{Actual} & Projected  & \multicolumn{2}{c}{Actual} & Projected  & \multicolumn{2}{c}{Actual} \\ \hline
Short (20 mm)                  & 21.52     & 21.82     & $\pm$ 0.02     & 21.52      & 21.73     & $\pm$ 0.02     & 20         & 17.23     & $\pm$ 0.02     \\
Medium (50 mm)                 & 21.52     & 21.78     & $\pm$ 0.02     & 21.52      & 21.78     & $\pm$ 0.02     & 50         & 47.09     & $\pm$ 0.02     \\
Long (70 mm)                   & 21.52     & 21.87     & $\pm$ 0.02     & 21.52      & 21.92     & $\pm$ 0.02     & 70         & 68.08     & $\pm$ 0.02     \\ \hline
\end{tabular}%
}
\end{table}

\Cref{ArrayDim} lists the final measured dimensions for the 2D arrays. For each array, the width given in the first column of \Cref{ArrayDim} represents the printed dimension of the 1D arrays. The second listed dimension (height) reflects direction of the machined face used for manual layer alignment. The pixel widths and heights are projected to be 21.52 mm $\times$ 21.52 mm when including the thicknesses of the seven 3-mm-pixels as well as the eight ESR foils (approximately 65 $\mu$m each). The active fraction of the array is 95.2\%. Most individual pixels demonstrated a deviation of less than 0.1 mm from the expected 3 mm measured across the array face. However, pixels along the outer edge of the array displayed a deviation as large as 0.3 mm due to the final polishing of their exterior surfaces. Actual measurements for the complete array averaged over five points showed the original 70 mm-long arrays to be 21.87 $\times$ 21.92 $\times$ 68.08 $\text{mm}^3$. The slight increase in width and height relative to the nominal design is attributed to excess resin at interlayer bond lines. The reduced length arises from material loss during sectioning and subsequent polishing. Sectioning with a 0.6 mm thick blade removes approximately 1.2 mm of material when both faces are cut, with additional losses arising from cut alignment tolerances and polishing required to achieve parallel, optically transparent faces. The 70 mm array was sectioned into 50 and 20 mm lengths, repolished, and remeasured with the results also shown in \Cref{ArrayDim}.

The pixel alignment along the machined face direction varied across the arrays due to the manual alignment and bonding. Several regions exhibited excellent alignment with negligible offset between adjacent layers, while other regions showed lateral shifts up to 0.4 mm when measured along the polished face, as illustrated in \Cref{2DArray}. This variation reflects the practical limits of manual stacking and alignment of individual pixel sections.

\subsection{Light Characterization}
\subsubsection{Gain Matching \& LO}

\begin{figure}[ht!]
    \centering
    \includegraphics[width=0.6\linewidth]{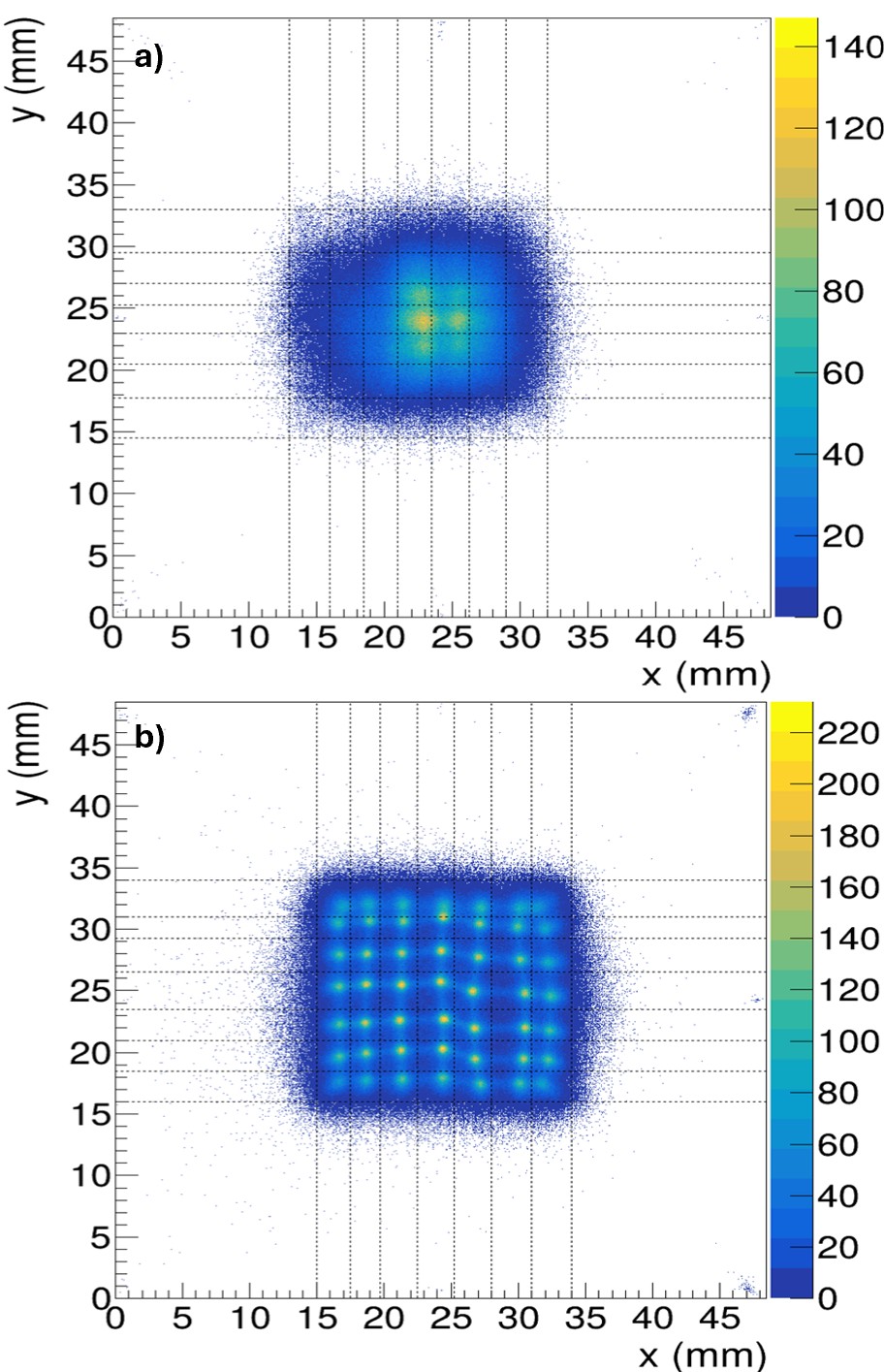}
    \caption{The a) 70 mm and b) 20 mm pixel array's XY position response when irradiated with a $^{207}$Bi source.}
    \label{Bi207XY}
\end{figure}

The initial 70 mm array was measured over 3 hours at 1'' distance inside the dark box assembly shown in \Cref{ArrayBox} with a 5 $\mu$Ci $^{207}$Bi radiation source, and the resulting XY position data calculated with \cref{XYEQ} is presented in \Cref{Bi207XY}a. Dashed lines indicate the estimated positions and dimensions of the individual pixels. While the central pixels are resolved, the outer pixels exhibit poor resolution and low statistics. One explanation for this reduction in performance is the incomplete coverage of the pixels with ESR film. ESR layers internal to the array are applied prior to final polishing, allowing the array ends to be polished to match dimensionally. In contrast, ESR is applied to external pixel array surfaces during post-processing. This means they must be aligned and matched as close as possible to the final dimensions as additional polishing risks delamination. Any slight mismatch introduced during this manual fitting can allow interactions to escape that would otherwise be captured. However, the most significant loss in light output and resolution stems from the proportions of the scintillator pixels. The 70 mm length and 3 mm width and height of a single pixel yield an aspect ratio exceeding 23. This large aspect ratio increases the number of internal reflections required for photons to reach the detector. Although the 3M ESR specular reflector exhibits very high reflectivity, repeated reflections lead to cumulative losses due to finite reflectivity and angular escape at the reflector interface \cite{Janecek2012}. Using a conservative reflectivity value of 97\%, approximately 50\% of the light is lost after ~23 reflections. Bulk absorption within the scintillator contributes to the overall loss but is secondary compared to interface-related losses for these high–aspect-ratio pixels. To address this, the significantly shorter 20 mm pixel array was measured under identical conditions: the same 3-hour duration, $^{207}$Bi source, and 1'' measurement distance. \Cref{Bi207XY}b shows that the shorter array exhibits significantly improved XY resolution, with all 49 pixel centers resolved. Nevertheless, the corner pixels still suffer from diminished resolution and light output compared to the center region. Additional blurring is still observed along the edges of the array due to degraded ESR adhesion or slight misalignment in pixel placement or size.

\begin{figure}[ht!]
    \centering
    \includegraphics[width=0.7\linewidth]{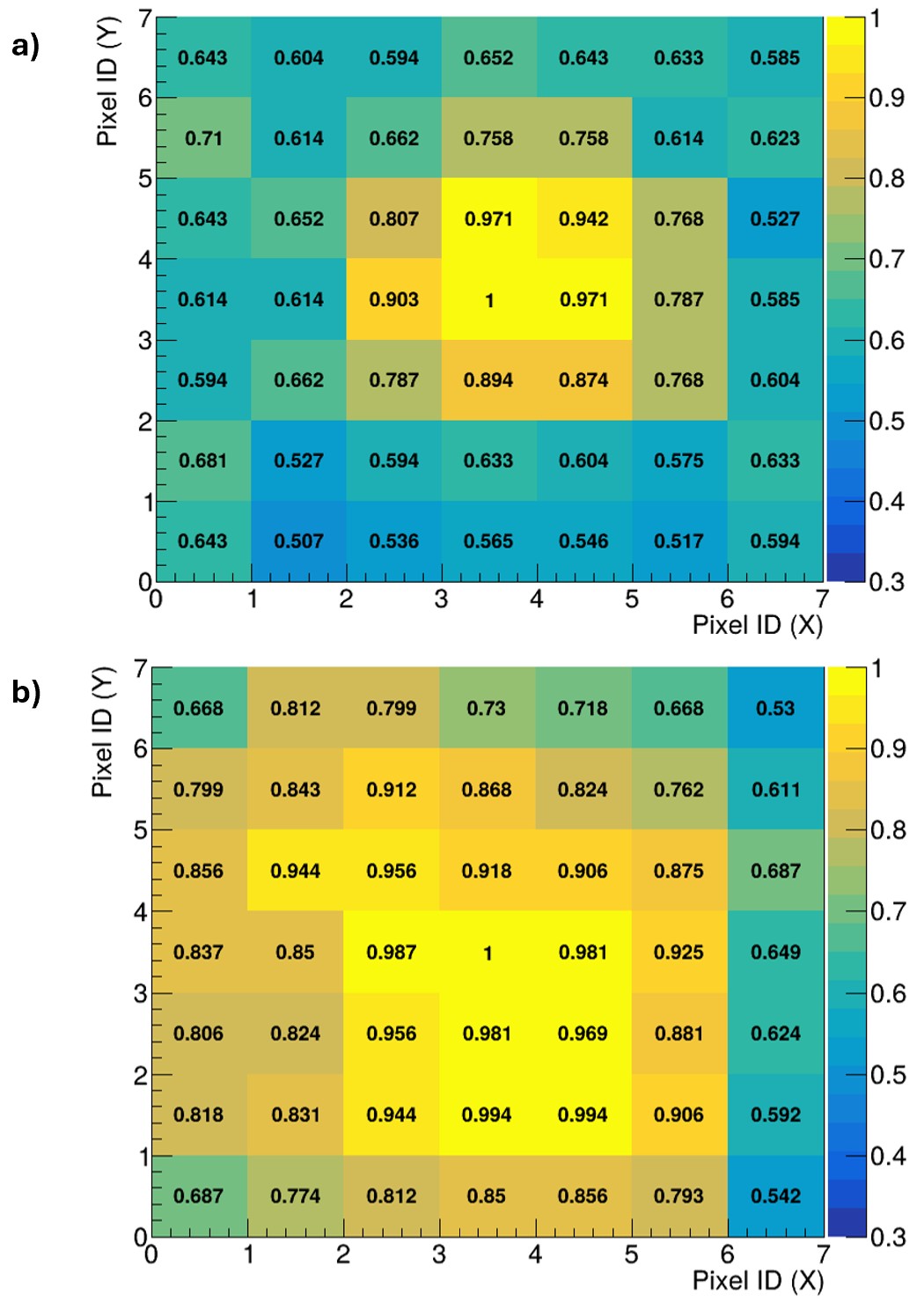}
    \caption{The a) 70 mm and b) 20 mm pixel array's gain variation for each pixel when irradiated with a $^{207}$Bi source and normalized to the respective center pixel (3,3) of each array. Prior to normalization, the center pixel of the 70 mm array exhibited a light output 0.291 times that of the 20 mm array center pixel under identical measurement conditions.}
    \label{Gainmap}
\end{figure}

The light output from each pixel is also shown to vary despite the nearly identical physical dimensions. In these arrays, the observed variation likely arises from a combination of fabrication and optical transport effects, including differences in surface quality, the quality of ESR bonding, dye uniformity in material composition, and light sharing between pixels. The bin corresponding to the $^{207}$Bi conversion electron peak for each pixel was identified and compared against a baseline pixel located in the center of the array. Following, a gain correction factor was computed as the ratio of the peak channel position of each pixel to that of this reference pixel. Prior to this internal normalization, the light output of the 70 mm array center pixel was 0.291 times that of the 20 mm array center pixel, indicating a substantially lower intrinsic light collection efficiency for the longer pixel geometry. The resulting relative gain factors are shown in \Cref{Gainmap}a and \Cref{Gainmap}b for the 70 mm and 20 mm array, respectively. The 20 mm array exhibits a more uniform response across pixels, whereas the 70 mm array shows a more pronounced decrease in light output toward the edges. This behavior is consistent with increased optical light sharing in higher aspect-ratio pixels, where the longer scintillation photon path length leads to a greater number of reflections at the pixel–reflector interfaces and an increased probability of photon loss or transfer into neighboring pixels. These effects are most pronounced for edge pixels, which experience reduced optical confinement and a higher likelihood of photon escape compared to centrally located pixels \cite{Knoll, Newby, Janecek2012}. In addition, the finite source distance and resulting angular distribution of incident radiation may contribute to variations in interaction depth and lateral light spread across the array, potentially enhancing edge nonuniformities. However, for array characterization this geometric effect is expected to be secondary compared to the dominant optical transport effects associated with pixel aspect ratio and array size.

\begin{figure}[ht!]
    \centering
    \includegraphics[width=0.55\linewidth]{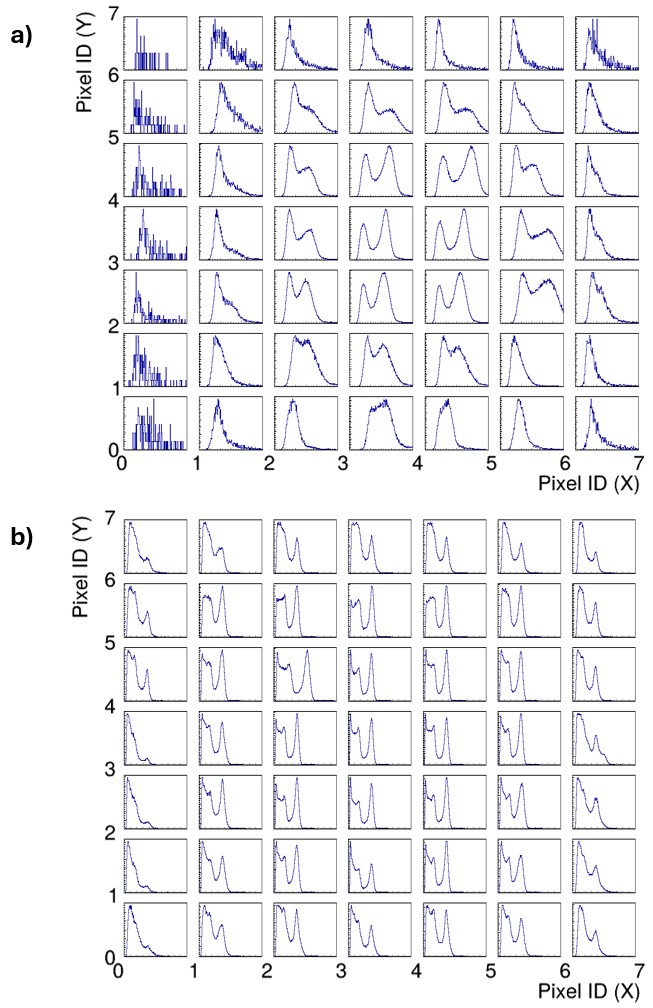}
    \caption{Individual pulse integral spectra histograms for a) 70 mm and b) 20 mm pixel arrays when irradiated with a $^{207}$Bi source.}
    \label{EHistPixel}
\end{figure}



The resulting ratio was then used to adjust the corresponding pixel histogram, shifting the peak to align with the reference pixel. \Cref{EHistPixel}a displays the light histograms for the 70 mm array, while \Cref{EHistPixel}b shows the gain-matched histograms for the 20 mm array. For the 70 mm array, center pixels exhibit sufficient light output and event statistics to resolve the conversion electron peak, making gain matching feasible. In contrast, the edge pixels show substantially reduced counts and poorly resolved peaks. This reduction arises from a combination of factors: photons generated in exterior pixels are more likely to escape the pixel face, slightly smaller pixel dimensions at the edges due to material loss during cutting and polishing, and local variations in ESR wrapping or adhesion. Together, these effects reduce the detected light for outer pixels, leading to the reduced count histograms observed in \Cref{EHistPixel}a. The 20 mm array, having a lower aspect ratio, exhibits sufficient light output and distinct conversion electron peaks across all 49 pixels, enabling straightforward gain matching.

 


\subsubsection{2D Array PSD Performance}

\begin{figure}[ht!]
    \centering
    \includegraphics[width=0.6\linewidth]{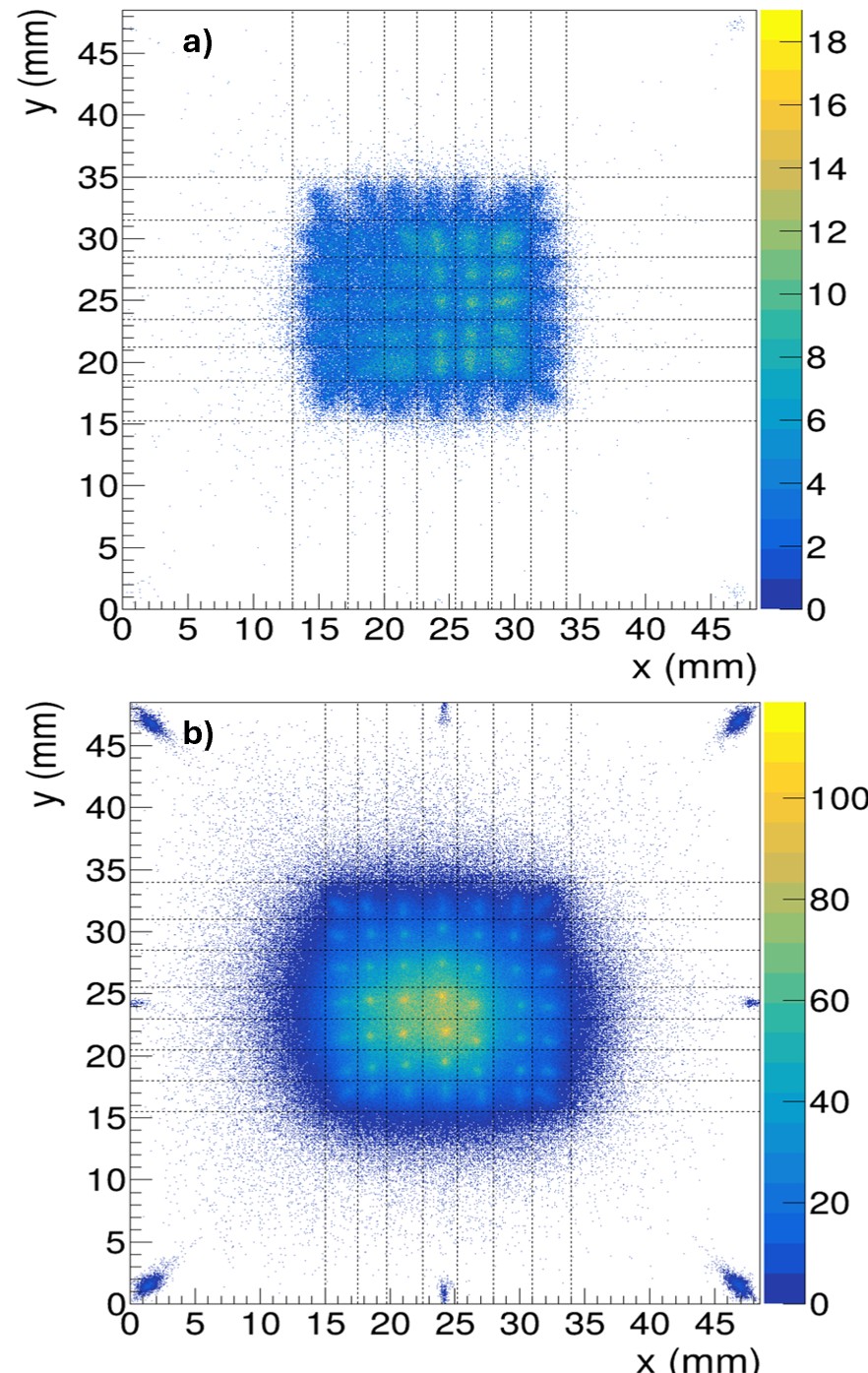}
    \caption{The a) 70 mm and b) 20 mm pixel array's XY position response when irradiated with an AmBe source.}
    \label{AmBeXY}
\end{figure}
 



The 70 mm and 20 mm array were evaluated for particle discrimination using an AmBe source. The resulting position reconstruction plots are shown in \Cref{AmBeXY}, with dashed lines indicating the estimated pixel locations and dimensions. The shorter array exhibits a noticeable improvement in resolution. This is primarily due to its significantly lower aspect ratio. Nevertheless, both arrays display distinguishable pixel structures. Consistent with the $^{207}$Bi results, the corner pixels show reduced resolution and light output due to incomplete ESR coverage or less surrounding material compared to the central pixels.

\begin{figure}[ht!]
    \centering
    \includegraphics[width=0.55\linewidth]{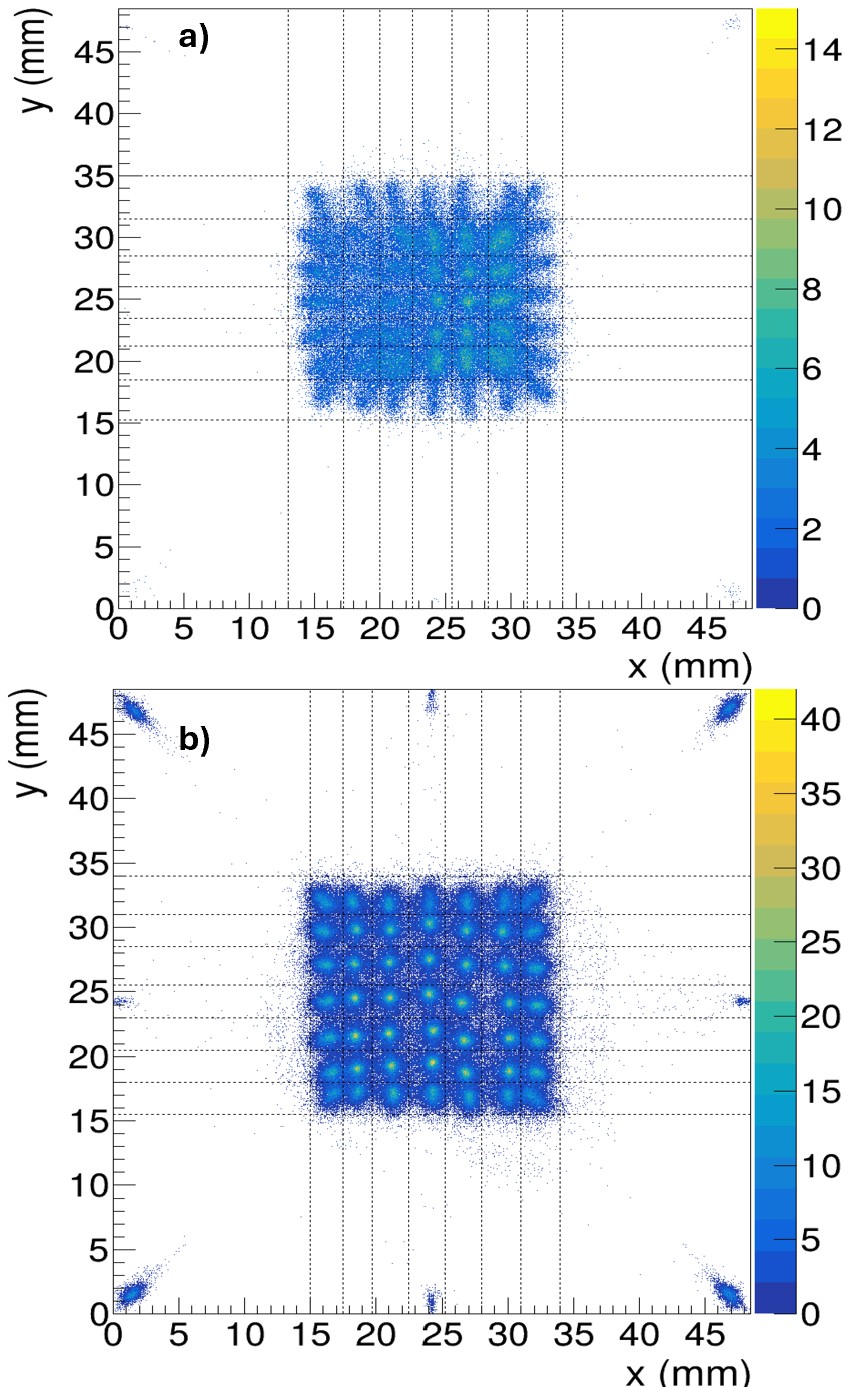}
    \caption{The a) 70 mm and b) 20 mm pixel array's XY position response when irradiated with an AmBe source above a low light threshold.}
    \label{AmBe20XY}
\end{figure}
 
\Cref{AmBe20XY} shows the XY position plot with a 0.2 MeV electron equivalent (MeVee) threshold for the 70 and 20 mm array to exclude low-light events. All 49 individual pixels are visible in both with improved distinction in the 20 mm array.



\begin{figure}[ht!]
    \centering
    \includegraphics[width=0.5\linewidth]{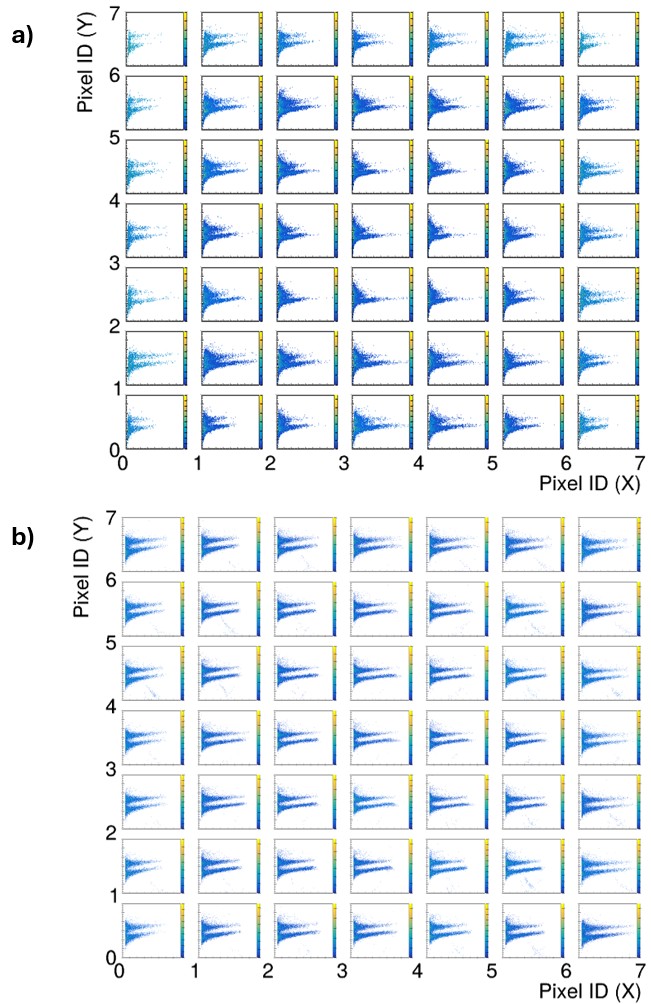}
    \caption{The a) 70 mm and b) 20 mm pixel array's individual pixel 2D PSD histogram when irradiated with an AmBe source.}
    \label{Individual2DPSD}
\end{figure}


\Cref{Individual2DPSD}a and b show the 2D PSD histograms for the 70 mm and 20 mm arrays, respectively, when PSD, defined in \cref{PSD}, is plotted individually for each pixel. PSD is discernible in both cases, but the outer edge pixels exhibit significantly lower statistics compared to the central pixels. As observed in \Cref{EHistPixel} the 70 mm array shows noticeably poorer particle-type separation relative to the 20 mm array. This degradation in resolution is present across all sides but is most pronounced on the far left side of the array. This suggests that this region may be the least uniform in terms of geometry due to inconsistent ESR adhesion or uneven polishing. Overall, particle discrimination is possible in every pixel for both arrays. The FoM for each pixel was evaluated for a light-output bin of 0.4–0.6 MeVee. For the 20 mm array, FoM values range between 1.1 and 1.3 for the central 5 $\times$ 5 pixels and remain above 0.84 for the exterior ring, yielding an average FoM of 1.06 with a standard deviation of 0.10 across all pixels. For the 70 mm array, FoM values range between 0.41 and 0.80, with most pixels exceeding 0.5 and an overall average FoM of 0.65 with a standard deviation of 0.11 across the pixels. The higher FoM values observed in the 20 mm array are consistent with the increased light output shown previously. As seen in \Cref{Individual2DPSD}, both array lengths demonstrate measurable neutron–gamma separation, with degradation at lower light-output regions and along the array edges, most pronounced in the 70 mm array.




\section{Conclusion}
\label{sec:conclusion}
In this work, we developed fabrication assembly tools and procedures for 2D pixelated plastic scintillator arrays. The scintillator is made from a non-aromatic acrylate-oligomer-base photocurable resin. The 2D arrays were manufactured in two main steps: a fully autonomous 1D array layering followed by a semi-autonomous cutting and pixel conversion step. The first step allowed fast manufacture of 1D arrays up to 9 layers thick and tight dimensional tolerances at a rate of around 4 layers per hour. Semi-manual conversion of 1D assemblies to 2D-pixel arrays allowed the production of arrays up to 7 $\times$ 7 and 70 mm long in around 3.5 hours. A complete 7 $\times$ 7 2D array ready for use can be manufactured from liquid resin to finished array in under 8.5 hours, including post processing. The primary limitation for the size of the arrays is the volume of the vat for resin and time to align and manually bond pixel layers together. 

Purple discoloration was present from the interaction of PPO and TPO in both 1D and 2D array manufacturing but quickly dissipated over 48 hours. Pixel yellowing was observed in manufacturing 2D arrays due to the repeated exposure to curing light and heat generated within the samples. This yellowing indicated degradation of the chromophores in the arrays and a reduction in light output. Other defects in the production of 2D-pixel arrays included bubbles formed within slabs from air becoming trapped during curing and leaching of fluors over time. Both defects were minimized with changes to the manufacturing process. Modifications included changes to how the build plate entered the resin vat, rest periods to allow cooling and air to escape the resin solution, and alcohol washes to expedite the leaching process. Despite the entirely autonomous nature of the 1D array manufacturing and the slow process of 2D array alignment, the final produced arrays showed tight dimensional tolerances. While designed arrays were expected to be 21.585 $\times$ 21.585 $\text{mm}^2$ in pixel area, produced arrays varied by less than 0.5 mm, with the difference coming from the variable nature of manually applying resin to bond the pixel sections together. The 2D arrays were measured at the full 70 mm length and then cut down to 20 mm for additional testing. All pixels in the short array exhibited sufficient resolution to discriminate neutron and gamma interactions and to identify conversion electron peaks. In contrast, the 70 mm array showed worse results due to a larger aspect ratio. The scintillation and resolution performance is improved in the 20 mm array while overall efficiency is greater within the 70 mm array. This illustrates the tradeoff between the length (i.e. efficiency) and scintillation performance of the arrays. 


The photocurable additive manufacturing approach demonstrated here exhibits reduced scintillation and PSD performance compared to a state-of-the-art (SOTA) pixelated plastic scintillator arrays produced via thermal polymerization and subtractive manufacturing. This performance gap is in part attributable to the non-aromatic acrylate-based resin, which is selected to enable rapid photopolymerization but inherently limits scintillation efficiency. The finer 3 mm pixel pitch of the additively manufactured arrays also results in a larger aspect ratio and therefore additional optical losses as light propagates through the pixel. Nevertheless, the developed approach offers several advantages relative to the SOTA, including reduced manufacturing time, lower manual labor requirements, and increased flexibility in array geometry. In addition, the layer-wise fabrication and resin bonding process eliminates the need for mechanical sectioning and polishing of thin scintillator bars, which are common failure points in conventional fabrication. As a result, this approach has the potential to enable finer pixel pitches that would be difficult to achieve using traditional machining-based methods. However, the lower light output and PSD performance compared to the SOTA imply the need for a higher light threshold for the 3D-printed arrays, which worsens detection efficiency. These tradeoffs should be evaluated through experimental studies to assess their impact on imaging performance.

In parallel, automation of the subtractive manufacturing pathway warrants full investigation, particularly with respect to achievable accuracy, throughput, cost, and material waste. These complexities can be addressed in future work. Further improvements to the additive manufacturing route are also possible. One major opportunity is automating the 1D-to-2D array conversion, which is currently the slowest and least precise step. Work at Oak Ridge National Laboratory demonstrated that using a second 4-axis robotic arm to autonomously stack and align resin slabs and ESR with bonding resin can significantly accelerate this step. This can potentially reduce total array production time by up to two hours. Additional fabrication refinements aimed at minimizing air entrainment and thermal buildup may further improve array quality. Beyond process-level improvements, future work should also evaluate the scalability of this additive manufacturing approach, including its viability for high-throughput production, reproducibility across larger batch sizes, and integration into more fully industrialized workflows.

\section*{\raggedright \textbf{CRediT authorship contribution statement}}
Chandler Moore: Conceptualization, Methodology, Software, Validation, Investigation, Writing – original draft, Visualization. Juan Manfredi: Conceptualization, Methodology, Writing - Review \& Editing. Michael Febbraro: Conceptualization, Methodology, Writing - Review \& Editing. Daniel Rutstrom: Validation, Visualization, Writing - Review \& Editing. Andrew Decker: Conceptualization. Ryan Kemnitz: Resources. Thomas Ruland: Investigation. Paul Hausladen: Conceptualization, Writing - Review \& Editing.

\section*{\raggedright \textbf{Declaration of competing interest}}
The authors declare that they have no known competing financial interests or personal relationships that could have appeared to influence the work reported in this paper.
\section*{\raggedright \textbf{Data availability}}
Data will be made available on request.

\section*{\raggedright \textbf{Acknowledgments}}
This material is based upon work supported by the Department of Energy National Nuclear Security Administration through the Nuclear Science and Security Consortium under Award Number(s) DE-NA0003996 and the Defense Threat Reduction Agency under grant HDTRA1136911. The work was performed under the auspices of the U.S. Department of Energy by Oak Ridge National Laboratory under Contract DE-AC05-00OR22725 with UT-Battelle, LLC. This work has been supported by the U.S. Department of Energy office of Nonproliferation Research and Development (NA-22 Office).

\section*{\raggedright \textbf{Disclaimer}}

The views expressed in this document are those of the author and do not reflect the official policy or position of the United States Air Force, the United States Department of Defense, or the United States Government. This material is declared a work of the U.S. Government and is not subject to copyright protection in the United States. Imagery in this document is the property of the U.S. Air Force.





\bibliographystyle{elsarticle-num}
\bibliography{tns.bib}

\end{document}